\documentclass[12pt]{article}

\input epsf
\def\hybrid{\topmargin 0pt      \oddsidemargin 0pt
        \headheight 0pt \headsep 0pt
        \voffset=-0.5cm
        \textwidth 6.25in       
        \textheight 9.5in       
        \marginparwidth 0.0in
        \parskip 5pt plus 1pt   \jot = 1.5ex}
\catcode`\@=11
\def\marginnote#1{}

\newcount\hour
\newcount\minute
\newtoks\amorpm
\hour=\time\divide\hour by60
\minute=\time{\multiply\hour by60 \global\advance\minute by-\hour}
\edef\standardtime{{\ifnum\hour<12 \global\amorpm={am}%
        \else\global\amorpm={pm}\advance\hour by-12 \fi
        \ifnum\hour=0 \hour=12 \fi
        \number\hour:\ifnum\minute<10 0\fi\number\minute\the\amorpm}}
\edef\militarytime{\number\hour:\ifnum\minute<10 0\fi\number\minute}

\def\draftlabel#1{{\@bsphack\if@filesw {\let\thepage\relax
   \xdef\@gtempa{\write\@auxout{\string
      \newlabel{#1}{{\@currentlabel}{\thepage}}}}}\@gtempa
   \if@nobreak \ifvmode\nobreak\fi\fi\fi\@esphack}
        \gdef\@eqnlabel{#1}}
\def\@eqnlabel{}
\def\@vacuum{}
\def\draftmarginnote#1{\marginpar{\raggedright\scriptsize\tt#1}}
\def\draftlabel#1{{\@bsphack\if@filesw {\let\thepage\relax
   \xdef\@gtempa{\write\@auxout{\string
      \newlabel{#1}{{\@currentlabel}{\thepage}}}}}\@gtempa
   \if@nobreak \ifvmode\nobreak\fi\fi\fi\@esphack}
        \gdef\@eqnlabel{#1}}
\def\@eqnlabel{}
\def\@vacuum{}
\def\draftmarginnote#1{\marginpar{\raggedright\scriptsize\tt#1}}

\def\draft{\oddsidemargin -.5truein
        \def\@oddfoot{\sl preliminary draft \hfil
        \rm\thepage\hfil\sl\today\quad\militarytime}
        \let\@evenfoot\@oddfoot \overfullrule 3pt
        \let\label=\draftlabel
        \let\marginnote=\draftmarginnote
   \def\@eqnnum{(\theequation)\rlap{\kern\marginparsep\tt\@eqnlabel}%
\global\let\@eqnlabel\@vacuum}  }

\def\numberbysection{\@addtoreset{equation}{section}
        \def\theequation{\thesection.\arabic{equation}}}

\def\underline#1{\relax\ifmmode\@@underline#1\else
        $\@@underline{\hbox{#1}}$\relax\fi}

\def\titlepage{\@restonecolfalse\if@twocolumn\@restonecoltrue\onecolumn
     \else \newpage \fi \thispagestyle{empty}\c@page\z@
        \def\thefootnote{\fnsymbol{footnote}} }

\def\endtitlepage{\if@restonecol\twocolumn \else  \fi
        \def\thefootnote{\arabic{footnote}}
        \setcounter{footnote}{0}}  
\relax

\numberbysection
\hybrid

\newfont{\Bbb}{msbm10 scaled 1\@ptsize00}
\newcommand{\CC}{\mbox{\Bbb C}}

\newfont{\Bbbb}{msbm7 scaled 1\@ptsize00}
\newcommand{\z}{\raise-1pt\hbox{$\mbox{\Bbbb Z}$}}

\def\beq{\begin{equation}}
\def\eeq{\end{equation}}
\def\p{\partial}

\def\MM{{\sf M}}
\def\DD{{\sf D}}
\def\Dc{\CC \setminus {\sf D}}
\def\Uc{{\sf U}_{{\rm ext}}}
\def\UU{{\sf U}}
\def\BB{{\sf B}}

\def\lbracket{\left <}
\def\rbracket{\right >}
\def\vphcl{\varphi _{0}}
\def\rhocl{\rho _{0}}

\begin{document}

\begin{titlepage}

\title{Large $N$ expansion
for the 2D Dyson gas}

\author{A. Zabrodin
\thanks{Institute of Biochemical Physics,
Kosygina str. 4, 119991 Moscow, Russia
and ITEP, Bol. Cheremushkinskaya str. 25, 117259 Moscow, Russia}
\and
P. Wiegmann
\thanks{James Franck Institute and Enrico Fermi
Institute
of the University of Chicago, 5640 S.Ellis Avenue,
Chicago, IL 60637, USA and
Landau Institute for Theoretical Physics, Moscow, Russia}}

\date{December 2005}
\maketitle

\begin{abstract}

We discuss
the $1/N$ expansion
of the free energy
of $N$ logarithmically interacting charges in the plane
in an external field. For some particular values
of the inverse temperature $\beta$ this system
is equivalent to the eigenvalue version of
certain random matrix models, where it is referred to
as the ``Dyson gas" of eigenvalues.
To find the free energy at large $N$
and the structure of $1/N$-corrections, we first use
the effective action approach and then
confirm the results by solving the loop equation.
The results obtained give some new
representations of the mathematical objects
related to the Dirichlet boundary value problem,
complex analysis and spectral geometry of exterior
domains. They also suggest
interesting links with bosonic field theory
on Riemann surfaces, gravitational
anomalies and topological field theories.

\end{abstract}

\vfill

\end{titlepage}

\section{Introduction}

In this paper we discuss the large $N$ expansion
of the $N$-fold integral
\beq\label{I1}
Z_N =\int
|\Delta_N (z_i)|^{2\beta} \prod_{j=1}^{N}
e^{\frac{1}{\hbar}W(z_j )}d^2 z_j.
\eeq
Here
$W$ is a function of $z=x+iy$ and $\bar z =x-iy$,
$\beta >0$ is a parameter,
 $\Delta_N (z_i) =\prod_{i>j}^N (z_i -z_j )$
is the Vandermonde determinant, and $d^2 z \equiv dx \, dy$.
The ``Planck constant'' $\hbar$ is introduced here
to stress the quasiclassical nature of the large $N$ limit:
$N \to \infty$ together with $\hbar \to 0$ so that
$t_0 =N\hbar$ is kept fixed.

The integral is equal to the partition function of
the statistical ensemble of $N$
2D Coulomb charges in the external potential $W$
(the ``Dyson gas'' \cite{Dyson}).
Different aspects of 2D Coulomb plasma were discussed
in \cite{Jancovici}-\cite{Hastings},
for further applications see
\cite{KKMWZ}-\cite{TBAZW}.
In this interpretation, the parameter $\beta$ is the inverse temperature.
At large $\beta$
the Dyson gas is believed to form a Wigner crystal.
In this paper we assume, however,
that $\beta$ is such that the system
is in the liquid phase as $N\to \infty$.
At some particular values of $\beta$ the model can be
viewed as the eigenvalue version
of certain ensembles of random matrices \cite{Mehta}
(normal or complex matrices at
$\beta =1$ \cite{Zabor,GG} and normal self-dual matrices at $\beta =2$).

An important information is encoded in the
$1/N$-expansion
of the model.
For random matrices, it has the meaning
of the ``genus expansion"
since the $1/N$ order of perturbation theory
graphs is determined by their
Euler characteristics.
In the sequel, we prefer to work with the
equivalent $\hbar$-expansion, thus emphasizing
its semiclassical nature:
\beq\label{I2}
\log Z_N = c (N) + \frac{F_0}{\hbar^2} +
\frac{F_{1/2}}{\hbar} + F_1 + O(\hbar)
\eeq
The explicit form of the
$c(N)$ is given below.
It
can be absorbed by normalization.
With some abuse of terminology  we call
$\hbar^2\log Z_N$ a free energy.

When $N$ becomes large
new macroscopic structures emerge.
The gas segregates into ``phases''
with zero and non-zero density separated by a very
narrow interface. Let $\DD$ be the domain in the complex plane
where the density is non-zero
(it may consist of several disconnected
components). In the first approximation, the gas
looks like a continuous charged fluid trapped
in the domain $\DD$.
The density at any point outside it is exponentially
small as $N\to \infty$.

The first two terms in (\ref{I2}), $F_0$ and $F_{1/2}$,
are of purely classical nature in the sense that
only the static equilibrium state of the charges
(the saddle point of the integral) contributes to them.
The leading contribution to the free energy, $F_0$,
is basically the
Coulomb energy of the charged fluid in the domain $\DD$.
Taking into account
the discrete ``atomic" structure of the Dyson gas,
which implies a short-distance cutoff and entropy
of macroscopic states,
one is able to find the correction $F_{1/2}$ to
the ``classical'' free energy.
The next term, $F_1$, apart from further
corrections of the classical nature, includes contribution
from small fluctuations about the equilibrium state.

The $\beta =1$ Dyson gas confined to the line
is related to the model of Hermitian random matrices.
The $1/N$-expansion of this model beyond
the leading order
has been obtained in the seminal paper \cite{herm1}.
Recently, there was a progress
in understanding these results from the algebro-geometric
point of view \cite{Chekhov,Eynard2,CMMV}
and in extending them to other matrix
models \cite{Eynard1,Korotkin,WZles}.
In \cite{Kostov1}, the genus-1 correction was interpreted
in terms of free bosons on
Riemann surfaces.
Our results for $F_1$ (partially
reported in \cite{WZles}) enjoy a similar
interpretation and lead to interesting connections with
spectral geometry of planar domains.
In particular, our results suggest a formula
for the determinant of the Laplace operator
in exterior planar domains in terms of the conformal map
of the domain onto the exterior of the unit circle.
For polynomial potentials, $F_1$ enjoys
a finite determinant representation (\ref{det5})
similar to the one known in topological field
theories.

Our results for $F_1$ suggest a new deep
connection between the 2D Dyson gas and 2D quantum gravity.
This connection does not explore the well
known approach to random surfaces
through a scaling limit of
random matrices.  It rather indicates that the density of
2D Dyson particles can be treated
as a fluctuating 2D metric. We do not develop this approach in this paper.

In the rest of the introductory section
we fix the notation
and present some standard
exact relations to be used in the sequel.
We follow \cite{WZles,WZ2,Zles}.

The main observables in the Dyson gas statistical
ensemble are mean values and correlators
of symmetric functions of the particles coordinates.
Let $A(z_1 , \ldots , z_N)$ be such a function, then
the mean value $\lbracket A \rbracket $ is defined by
$$
\lbracket \, A \, \rbracket =\frac{1}{Z_N}
\int |\Delta_N (z_i)|^{2\beta}
A (z_1 , \ldots , z_N)
 \prod_{j=1}^{N} e^{\frac{1}{\hbar}W(z_j)}d^2 z_j
$$
A particularly important example is the density
\beq\label{density1}
\rho (z)=\hbar \sum_j \delta (z-z_j)
\eeq
where $\delta (z)$ is the two dimensional $\delta$-function.
Instead of correlations of density it is often convenient
to consider correlations of the field
\beq\label{potential1}
\varphi (z)
=-\beta \int \log |z-\zeta |^2 \rho (\zeta ) d^2 \zeta
\eeq
from which the correlations of density can be found
by means of the relation
\beq\label{rhophi}
4\pi \beta \rho (z)=-\Delta \varphi (z)
\eeq
Here and below, $\Delta =4 \p_z \p_{\bar z}$ is the Laplace operator.
Clearly, $\varphi$ is the 2D Coulomb potential
created by the charges.

Handling with multipoint correlation functions,
it is customary to pass to their connected parts.
For example, in the case of 2-point functions, the
connected correlation function is defined as
$$
\lbracket \rho (z_1 )\rho (z_2)\rbracket _{c}\equiv
\lbracket \rho (z_1 )\rho (z_2)\rbracket -
\lbracket \rho (z_1) \rbracket
\lbracket \rho (z_2) \rbracket
$$
The following variational formulas
hold true:
\beq\label{var}
\lbracket \rho (z) \rbracket =
\hbar^2 \, \frac{\delta \log Z_N}{\delta W(z)}\,,
\;\;\;\;
\lbracket \rho (z_1 )\rho (z_2)\rbracket _{c}=\hbar^2 \,
\frac{\delta \lbracket \rho (z_1)\rbracket }{\delta W(z_2)}=
\hbar^4 \frac{\delta^2 \log Z_N}{\delta W(z_1) \delta W(z_2)}
\eeq
These formulas are exact for any finite $N$.
They follow from the fact that
variation of the partition function over a general
potential $W$ inserts $\sum_i \delta (z-z_i)$ into
the integral.
More generally, the connected part of the
$(n+1)$-point density correlation function is given by the
linear response of the $n$-point one to a small variation
of the potential.

Let $f=f(z, \bar z)$ be a function in the complex plane
(for brevity we write simply $f(z)$ in what follows).
Summing over the charges, we get the symmetric function
$\sum_i f(z_i) \equiv {\mbox tr} f$, where the notation is
inspired by related models of random matrices.
Mean values and correlators of such functions are
expressed through those of densities:
\beq\label{tdens}
\lbracket \mbox{tr}\, f \rbracket =
\int \lbracket \rho (z) \rbracket f(z) d^2 z\,,
\quad
\lbracket \mbox{tr}\, f_1 \, \mbox{tr}\, f_2 \rbracket _{c}
=\int \lbracket \rho (z_1)\rho (z_2) \rbracket _{c}
f_1(z_1)f_2 (z_2) d^2 z_1 d^2 z_2
\eeq
and so on.

\section{Ward identities}

Clearly, the integral (\ref{I1}) remains the same
if we change the integration variables. In other words,
it is invariant under reparametrizations of the
$z$-coordinate.
This leads to a number of Ward identities which are
our basic tool to compute the free energy at large $N$.

\subsection{Holomorphic form of  the Ward identity
at finite $N$:  loop equation
\label{LL}}

We begin with a holomorphic reparametrization
$z_i \rightarrow z_i +\epsilon(z_i)$.
Let us apply it to  the integral (\ref{I1}),
$Z_N =\int e^{-\frac{1}{\hbar }E(z_1 , \ldots , z_N)}
\prod_j d^2 z_j$,
where the energy is
\beq\label{betaE}
-\hbar^{-1 }E =
\beta \sum_{i\neq j} \log |z_i - z_j | + \hbar^{-1}
\sum_j W(z_j)
\eeq
In the first order the integrand transforms as
$$
E\longrightarrow
E+ \sum_{l}\left ( \frac{\p E}{\p z_l}\, \epsilon (z_i) +
\frac{\p E}{\p \bar z_l}\, \overline{ \epsilon (z_i)}\right )
$$
while the volume element
$\prod_j d^2 z_j$ undergoes the scaling  (Weyl) transformation
$$
\prod_j d^2 z_j \longrightarrow
\left [ 1+ \sum_l (\p \epsilon (z_l) \! +\! \overline{ \p
 \epsilon (z_l)}\,)\right ]
\prod_j d^2 z_j
$$
The invariance of the integral is then expressed by
the identity
$$
\sum_i \int \frac{\p}{\p z_i} \left (
\epsilon (z_i) e^{-\frac{1}{\hbar }E}\right ) \prod_j d^2 z_j =0
$$
valid for any $\epsilon$. Introducing a suitable cutoff
at infinity, if necessary, one sees that
the 2D integral over $z_i$
can be transformed, by virtue of the Green theorem,
into a contour integral around infinity and so it does
vanish.

Let us take
\beq\label{e}\epsilon (z_i)=\frac{\epsilon}{z-z_i}\, ,
\eeq where
$z$ is a complex parameter. The singularity
at the point $z$ does not destroy the above identity
since its contribution is proportional to
the vanishing integral
$\oint d\bar z_i /(z_i -z)$ over a small contour
encircling $z$. We explore this singularity in the sequel.
Therefore, we have the equality
$$
\sum_i \int \left [
-\, \frac{ \p_{z_i}E}{z-z_i}+\frac{\hbar}{(z-z_i)^2}
\right ] e^{-\frac{1}{\hbar }E} \prod_j d^2 z_j =0
$$
where
$\displaystyle{\hbar^{-1}\p_{z_i}E=-
\beta\sum_{l\neq i}\frac{1}{z_i -z_l}
- \, \hbar^{-1} \p W(z_i)}$.
Using the identity
$$
\sum_{i,j}\frac{1}{(z-z_i)(z-z_j)}
=\sum_{i\neq j}\frac{2}{(z-z_i)(z_i -z_j)}+
\sum_i \frac{1}{(z-z_i)^2}
$$
we rewrite it in the form
\beq\label{T}
\lbracket {\cal T}\rbracket=0,
\eeq
where we define the holomorphic component of the stress energy tensor
\beq\label{TT}
-2\beta{\cal T}= \frac{2}{\hbar} \sum_i
\frac{\p W(z_i)}{z-z_i} \! +\! \beta
\left ( \sum_i \frac{1}{z\! -\! z_i}\right )^2 +
(2\!-\! \beta) \sum_i \frac{1}{(z-z_i)^2}
\eeq
This is the holomorphic form of the Ward identity.
In terms of the field
$\varphi (z)$ (\ref{potential1}) it reads
\beq\label{loopeq}
\frac{1}{2\pi}\int \frac{\p W(\zeta )
\lbracket \Delta \varphi (\zeta )\rbracket}{z-\zeta}
\, d^2 \zeta =
\lbracket (\p \varphi (z))^2 \rbracket
+(2\! -\! \beta )\hbar \, \lbracket \p^2 \varphi (z)\rbracket
\eeq
The correlator at coinciding points is understood as
$\displaystyle{\lbracket (\p \varphi (z))^2 \rbracket
=\lim_{z'\to z}
\lbracket \p \varphi (z) \, \p \varphi (z')\rbracket}$.

We have got an
exact relation
between one- and two-point correlation functions,
valid for any finite $N$. For historical reasons, it is
called the {\it loop equation}.
Since correlation functions are
variational derivatives of the free energy, the loop equation
is an implicit functional relation for the free energy.
However, it is not a closed relation.
It can be made closed by some additional assumptions
or approximations.
A combination with the $1/N$-expansion
is particularly meaningful.

\subsection{Path integral representation
of the partition function
and Weyl form of the Ward identity}

Here we give somewhat heuristic but
transparent arguments to calculate the free energy by
using the path integral representation and
the Weyl form of the Ward identity.
The results will be justified
and refined by means of the loop equation.

\paragraph{Density as a metric.}
At large $N$, when  the density  can be treated as a
smooth function, the partition function of
the Dyson gas can be represented as a path integral
over densities $\rho(z)$.
Symbolically, we write the partition function as
\beq\label{Z}
Z_N = \int [D\rho ] e^{- \, \frac{1}{\hbar^2} {\cal A}[\rho ]}
\eeq
where the action $\cal A
$ is to be determined.
In this approach $\rho (z)\, dzd\bar z $
appears as a metric $ g_{ab}d x^a dx^b$
written in the conformal gauge
$g_{ab}=\delta_{ab}\rho(z)$,
while the loop equation has the meaning
of the Ward identity with respect to
holomorphic diffeomorphisms.

\paragraph{Stress energy tensor.}
The action in Eq. (\ref{Z}) can be determined
by its response to a change of the metric.
The response of the action to variation of
the metric is generated by the stress energy tensor (s.e.t.).
Under the
Weyl transformation $\rho \to \rho+\delta \rho$,
the change of the action is
\beq\label{17}
-\delta {\cal A}=\frac{1}{\pi}\int
{\cal T}_{z\bar z}\, \delta \rho \,\rho^{-1} \, d^2z
\eeq
where
${\cal T}_{z\bar z}$
is the trace of the s.e.t.
The easiest way to
determine ${\cal T}_{z\bar z}$
is to use the conservation law of the s.e.t.
reflecting the reparametrization invariance of the action.
In the conformal gauge the conservation law reads
\beq\label{C}
\bar\p {\cal T}+
\rho \p (\rho^{-1} {\cal T}_{z\bar z})=0
\eeq
where ${\cal T}$
is the holomorphic component of the s.e.t.
It has been already
derived directly from the finite
dimensional integral (\ref{I1}). The result (Eq. (\ref{TT}))
can be rewritten as
\beq\label{T1}
-2\beta{\cal T}(z)=2\beta\int \frac{\p W(\zeta )}{z-\zeta}
\, \rho (\zeta ) d^2 \zeta +
(\p \varphi (z))^2 +(2\! -\! \beta )\hbar \, \p^2 \varphi (z)
\eeq
Let us apply $\bar \p$ to
this equality. Taking into account that
$\bar\p(1/z)=\pi\delta (z)$
and using (\ref{rhophi}), we obtain
$\bar \p{\cal T}= \pi \, \rho \left (\p \varphi - \p W +
 \frac{\hbar}{2} (2-\beta ) \p \log \rho \right )$.
The conservation law then states that
\beq\label{Tzbarz}
{\cal T}_{z\bar z}=\pi \rho \left [W-\varphi -
\frac{1}{2} (2-\beta )\hbar \log \left (e^{\lambda
+1} \rho \right )
\right ]
\eeq
where $\lambda$ is the integration constant which
will be determined by
the normalization condition
$\int \lbracket\rho\rbracket d^2z=\hbar N$.

\paragraph{The action.}
Integrating Eq. (\ref{17}) with
${\cal T}_{z\bar z}$ given by (\ref{Tzbarz}), one
determines
the action\footnote{The effective
action for some other
models of random matrices has been discussed in Refrs.
\cite{collective,herm-col}.} up to a constant:
\beq\label{A}
-{\cal A}=
\frac{1}{8\pi}\int \left (\beta^{-1} \varphi
(\rho^{-1}\Delta) \varphi +8\pi W -
4 \pi (2\! -\! \beta)\hbar \log(e^{\lambda }\rho )
\right )\rho \, d^2 z
\eeq
Here $\rho^{-1}\Delta$ is the invariant Laplace-Beltrami
operator in the metric $\rho\, dzd\bar z$.

Let us discuss the physical meaning
of this action.
It consists of energy and entropy contributions.
The energy is given by (\ref{betaE}).
On the scales much larger than
the mean distance between the charges
the system can be treated as
a charged liquid with the electrostatic energy
\beq\label{betaE1}
- \hbar E_0 [\rho ]=\beta
\int \!\! \int \rho (z)
\log |z-\zeta | \, \rho (z' ) d^2 z d^2 z'
\, +\, \int W(z)\rho (z) d^2 z
\eeq
$$=\frac{1}{8\pi}
\int \left (\beta^{-1} \varphi (\rho^{-1}\Delta)
\varphi +
8\pi W \right )\rho \, d^2 z \, .
$$
It gives the leading term of the action. The correction
$\frac{\hbar}{2}(2 \! -\! \beta ) \int \rho \log \rho \,
d^2 z$
results from the
discrete ``atomic" structure of the Dyson gas.
The argument below goes back to Dyson
\cite{Dyson}.

The subleading term of the action
consists of two contributions
of different nature.
One correction comes from the sum
$\sum_{i\neq j} \log |z_i - z_j |$, when
passing to the continuous theory. Namely, one should
exclude the terms with $i=j$, writing
$$
\sum_{i\neq j} \log |z_i - z_j | =
\sum_{i,j} \log |z_i - z_j |
-\sum_j \log \ell (z_j )
$$
or
$$
\sum_{i\neq j} \log |z_i - z_j | =
\hbar^{-2}\int \rho (z) \log |z-z'|\rho (z') d^2 z d^2 z'
-\hbar^{-1}\int \rho (z) \log \ell (z) \, d^2 z
$$
where $\ell$ is a short-distance cutoff
(which may depend on the point $z_j$). It is natural to
take the cutoff to be
\beq\label{cutoff}
\ell (z) \simeq \sqrt{ \frac{\hbar}{\rho (z)}}
\eeq
which is the mean distance between
the charges around the point $z$.
This gives the
improved estimate for the electrostatic energy:
\beq\label{smdist}
E[\rho ] = E_0[\rho] - \frac{1}{2}\beta
\int \rho  \log \rho  \, d^2 z +
\frac{1}{2}\beta\hbar N \log \hbar
-\gamma_1 \hbar N
\eeq
where $\gamma_1$ is a numerical constant which can not be determined
by this argument.

Another correction comes from the integration measure
when one passes from the integration over $z_j$ to the
integration over macroscopic densities. We can write
$$
\prod_j d^2 z_j = N! \, J[\rho ] \, [D\rho ]
$$
where $[D\rho ]$ is an integration measure in the space
of densities, $J[\rho ]$ is the Jacobian of this change of variables
and the factor $N!$ takes into account
the symmetry under permutations (all the states
that differ by a permutation of the charges
are identical). To estimate the Jacobian,
we divide the plane into $N$ microscopic ``cells" such that
$j$-th particle occupies a cell of size
$\ell (z_j)$, where
$\ell (z_j)$ is the mean distance (\ref{cutoff})
between the particles
around the point $z_j$.
All the microscopic states in which the particles
remain in their cells are macroscopically indistinguishable.
Given a macroscopic density $\rho$,
$J[\rho ]$ is then approximately equal to the integral
$\int_{{\rm cells}} \prod_j d^2 z_j$, with each particle
being confined to
its own cell. Therefore,
$J[\rho ] \sim \prod_j \ell^2 (z_j)$, and thus
$\log J[\rho ]$ (sometimes referred to as entropy
of the state with the macroscopic density $\rho$)
is given by
\beq\label{Jacob}
\hbar \log J[\rho ] = -\int \rho
\log \rho \, d^2 z \, + \, \hbar N \log \hbar +\gamma_2 \hbar N
\eeq
where $\gamma_2$ is another numerical
constant\footnote{
Combining (\ref{smdist}), (\ref{Jacob})
and taking into account the factor $N!$
in the measure, we obtain for the
$c(N)$ in Eq. (\ref{I2}):
$c(N)= \log N! + \frac{N}{2}(2-\beta ) \log \hbar
+\gamma N$,
where $\gamma =\gamma_1 +
\gamma_2$ is a numerical constant.}.

The subleading term of the action,
$ -\, \frac{1}{2}(2\! -\! \beta)\int
\rho \, \log \rho \, d^2 z$,
is thus the sum of the contribution
due to the short distance cutoff and the entropy contribution.
They cancel each other in the ensemble of normal self-dual
matrices (at $\beta =2$).

\paragraph{The Weyl form of the loop equation.}
The gravitational Ward identities
state that the expectation value of the variation of the action
vanishes.
They can  be written in two complimentary (holomorphic and Weyl) forms
\beq\label{W1}
\lbracket{\cal T}\rbracket=0,
\quad
\lbracket{\cal T}_{z\bar z}\rbracket=0\, .
\eeq
The first identity, generated by the holomorphic component of the
s.e.t.,  is the holomorphic loop equation (\ref{loopeq}).
The second one,
$
-\frac{1}{\pi}\lbracket {\cal T}_{z\bar z}
\rbracket = \lbracket \rho \frac{\delta {\cal A}}{\delta \rho}
\rbracket=0 \, ,
$
i.e.,
\beq\label{L}
\lbracket \varphi \, \rho \rbracket
-W\lbracket \rho \rbracket +
\frac{1}{2} (2\! -\! \beta ) \hbar
\lbracket \rho \log (e^{\lambda +1} \rho )\rbracket=0
\eeq
is another form of the loop equation.

The Weyl form of the Ward identity represents
the gravitational anomaly.
It is based on the fact that the diffeomorphism
(\ref{e}) is not holomorphic at the points $z_i$ (positions of the
particles). At these points $\bar \p
\epsilon (z)= \epsilon \pi \delta (z-z_i)$ and
the holomorphic component ${\cal T}$
of the s.e.t. is not analytic.
It has simple and double poles.

\section{The structure of the $1/N$ expansion}

The large $N$ limit we are interested in ($N\hbar =t_0$
remains finite) corresponds to
a very low effective temperature of the gas, when fluctuations
around equilibrium positions of the charges are
negligible. The main contribution to the partition function
then comes from a configuration,
where the charges are ``frozen" at their equilibrium
positions and, moreover, the gas can be treated as a
continuous fluid at static equilibrium.
Mathematically, all this means that the integral (\ref{I1}) is
evaluated by the saddle point method, with only the
leading contribution being taken into account.
Fluctuations around the saddle point give
$1/N$ corrections.

The path integral (\ref{Z})
makes the structure of the $1/N$
expansion intuitively clear. First we find
an equilibrium (``classical")
density $\rho_{cl}$ which minimizes the action
${\cal A}[\rho ]$.
Then, separating
the classical part of the density,
$\rho =\rho_{cl}+\hbar \delta\rho$ we can write
$$
{\cal A}[\rho]=
{\cal A}[\rho_{cl}]+\frac{\hbar^2}{2}\int
\delta\rho(z)\,
K(z,z') \delta\rho(z')d^2z+\dots
$$
where
$$
\left. K(z,z')=\frac{\delta^2 {\cal A}[\rho ]}{\delta
\rho (z) \delta \rho (z')}
 \right |_{\rho = \rho_{cl}}
$$
is the kernel of an integral operator $\hat K$.
The path integral representation
of the Dyson gas immediately produces
the first two leading contributions to the free energy:
\beq\label{F1}
\hbar^2\log Z_N=-{\cal A}[ \rho_{cl}] -
\frac{\hbar^{2}}{2}\log \mbox{det}
\hat K \, +O(\hbar^3 )
\eeq
The first term is the classical value
of the action, the second one is due to the Gaussian
fluctuations around the classical solution\footnote{One can
obtain  this results by a direct
iteration of the Ward identity (\ref{L})
without appealing to the path integral  representation.}.

Since the leading part of the free energy is the
classical value of the action,
the connected part of the
pair density correlation computed in the leading order
is equal to the kernel of the operator inverse to
$\hat K$:
\beq\label{c1}
\hbar^{-2}\lbracket \rho (z) \rho (z')\rbracket_c
=\lbracket \delta \rho (z)\, \delta \rho (z')\rbracket=
\hat K^{-1} (z, z')\quad \quad (\hbar \to 0)
\eeq
The term coming from
fluctuations (the second term in (\ref{F1}))
is not the only contribution to $F_1$.
 The action itself and the  solution $\varphi_{cl}$
of the equation
\beq\label{L1}
W-\varphi_{cl} =\frac{1}{2} (2 \! -\! \beta )
\hbar \log \left (-\frac{e^{\lambda +1}}{4\pi\beta}
 \Delta \varphi_{cl}\, \right )
\eeq
minimizing the action depend on $\hbar$.  Expanding the density,
$\rho_{cl}=\rho_0 +
\hbar \rho_{1/2} +\dots$,
we obtain
\beq\label{a}
{\cal A}[\rho_{cl}]={\cal A}[ \rho_0 ]-
\frac{\hbar^2}{2}\int \rho_{1/2}(z)K(z,z')\rho_{1/2}(z') d^2zd^2z'
+\dots,
\eeq
The action evaluated at $\rho_0$
yields the first two leading terms of the free energy.
Up to a constant
we write
$$
-{\cal A}[ \rho_0]=F_0+\hbar F_{1/2}\,.
$$
The second term in (\ref{a}) contributes to the next order:
$$
-{\cal A}[\rho_{cl}]+{\cal A}[ \rho_0]=\hbar^2 F_1^{(1)}\,.
$$

Summing up,
we obtain three leading orders of the  $1/N$ expansion
(\ref{I2}):
\beq\label{f0}
F_0 = - \beta\int\!\! \int
\rho_0 (z)\log \left | \frac{1}{z}\!-\! \frac{1}{\zeta}\right |
 \rho_0 (\zeta) d^2 z d^2 \zeta
\eeq
\beq\label{f1/2}
F_{1/2} = -\, \frac{1}{2}(2-\beta)\int
\rho_0 \log\rho_0 \, d^2 z
\eeq
\beq\label{F11}
F_1=F_1^{(0)}+F_1^{(1)}
\eeq
\beq\label{f1}
F_1^{(0)}=-\frac{1}{2}\log \mbox{det} \hat K,\eeq
\beq\label{f11}
 F_1^{(1)}=
\frac{1}{2}\int \rho_{1/2}(z)K(z,z')\rho_{1/2}(z') d^2zd^2z'
\eeq
They are expressed
entirely through the mean density and the pair
correlation function (\ref{c1}) computed
in the leading order.
This remains to be the case for the higher order corrections as well.

The mean density and the pair correlation function were found in
\cite{WZ2}.
Using these results one is able to compute
the free energy by means of (\ref{f0})--(\ref{f11}).
We summarize
the results in the next section. In the sequel we derive them on
a more solid basis
by a direct iteration  of the holomorphic loop equation.

We note also that the
$\hbar$-expansion of the free energy can be written in the
 ``topological" form
$\sum_{g\geq 0} \hbar^{2g}  \tilde F_g$, where each term has its
own expansion in  $\varepsilon =(2-\beta )\hbar$:
$ \tilde F_g =\tilde F_{g}^{(0)}+\sum_{n\geq 1}
\varepsilon^n \tilde F_{g}^{(n)}$.

\section{The main results\label{Su}}

We introduce the following notation:
\beq\label{sigma}
\sigma (z)=-\frac{1}{4\pi}\Delta W(z),\quad
\chi (z) =\log \sqrt{\pi \sigma (z)}, \quad
\alpha =\sqrt{\frac{2}{\beta}}-
\sqrt{\frac{\beta}{2}}
\eeq

\subsection{Summary of the results}

In order to compute the first three leading
contributions to the free energy we need the following
results.

\begin{itemize}
\item
{\bf The mean density} computed as a power expansion
in $\hbar$  vanishes outside a bounded domain $\DD$
(we assume that $\DD$ is connected).
Inside the domain
the first two orders are
\beq\label{d}
\rho_0=\beta^{-1}\sigma, \quad  z\in \DD;
\eeq
\beq\label{12}
\rho_{1/2}=-(2-\beta)\hat K^{-1}\chi\quad  z\in \DD .
\eeq
The  leading  correction to the density $\rho_{1/2}$ is
singular at the boundary (see Eq.(\ref{S2})).
Corrections exponential in $\hbar$   make the density
 a smooth function falling exponentially  outside
$\DD$.

\item
{\bf The shape of $\DD$} is
determined by the potential $W$ and the
number of particles, as is described below
(see (\ref{shape})).
It is the subject of the inverse potential problem
in 2D.
\item {\bf The pair correlation function of densities}
is related to the Dirichlet boundary problem in
the complimentary domain $\Dc$ (the exterior of
$\DD$).
Given a function $f(z)$,
let $f^H (z)$ be
its harmonic continuation from the boundary
to the exterior of $\DD$ (the
solution
of the exterior Dirichlet boundary value problem).
The pair correlation function of densities (and thus
the kernel $K(z_1 , z_2)$) is completely
characterized
by the integral relation
\beq\label{2trace11}
4\pi \beta \int f(z)\hat K^{-1}(z,z')\,g(z')d^2z d^2z'
=-\int_{\Dc} \nabla (f-f^H)\nabla (g-g^H)   d^2 z +
\int_{\CC} \nabla f \nabla g d^2 z
\eeq
for any smooth functions $f$, $g$.
Hereafter, $\hat K^{-1}(z,z')$ is the kernel of
the operator $\hat K^{-1}$.
The first
integral goes over the exterior of the
domain $\DD$ while
the second one is over the entire plane. Note
that $f-f^H$ vanishes on the boundary.
\item {\bf The spectral determinant
$\det \hat K$} in (\ref{f1})
is a ratio of two spectral determinants of the
invariant Laplace-Beltrami operators
$e^{-2\chi}\Delta$
in the conformal metric
$\pi \sigma =e^{2\chi}$:
\beq\label{K}
\log \det \hat K=\log \frac{\det (-e^{-2\chi}
\Delta_{\Dc})}{ \det (-e^{-2\chi}\Delta_{\CC})}.
\eeq
One of them acts on the entire plane (and so
does not depend on $\DD$).
The other one is the Laplace-Beltrami
operator in the {\it exterior domain} $\Dc$ acting
on functions vanishing at the boundary.
The determinants of the Laplacians in exterior,
unbounded domains were introduced in \cite{Zelditch}.
The definition is more involved than the usual
zeta-function regularization.
For the details see
\cite{Zelditch} and references therein.
However, explicit formulas for the exterior
determinants were not known.
Our result for $F_{1}^{(0)}$ obtained by
a direct solution of the iterated loop equation
suggests such a formula (see below).
\end{itemize}

To present the results for the free energy we need
some more notation.
Let $ds \equiv |dz|$ be the line element along the
boundary of $\DD$, $\kappa$ be the local curvature
of the boundary, and
$e^\phi dwd\bar w=dzd\bar z$ be the conformal metric induced
by the conformal map $w(z)$
of the exterior of $\DD$ onto the exterior of
the unit circle.

\begin{itemize}
\item {\bf Free energy:}
\beq\label{free1}
F_0 = - \frac{1}{\beta}\int_{\DD}\!\! \int_{\DD}
\sigma (z)\log \left | \frac{1}{z}\!-\! \frac{1}{\zeta}\right |
 \sigma (\zeta) d^2 z d^2 \zeta
\eeq
\beq\label{F1/2}
F_{1/2} = -\, \frac{2-\beta}{2\beta}\int_{\DD}
\sigma \log (\pi \sigma ) \, d^2 z
\eeq
\begin{eqnarray}\label{F10}
F_1^{(0)}& =&\displaystyle{
\frac{1}{24\pi} \left [
\int _{|w|>1} |\nabla (\phi +\chi ) |^2 \, d^2 w
-2\oint _{|w|=1} \! (\phi +\chi ) \, |dw| \right ]}
\nonumber\\
&&-\, \frac{1}{24\pi}\int_{\CC} |\nabla \chi |^2 \, d^2 z
- \frac{1}{8\pi}\oint_{\p \DD}\p_n \chi \, ds
\end{eqnarray}
\beq\label{F111}
F^{(1)}_1=-\frac{\alpha^2}{4\pi}
\left [\int_{\Dc} |\nabla (\chi-\chi^H)|^2  d^2 z -
\int_{\CC} |\nabla \chi|^2  d^2 z \right ]
+\frac{\mu + \frac{1}{4}}{2\pi}
\oint_{\p \DD}\p_n \chi \, ds
\eeq
Note that $F_{1/2}$ here differs from (\ref{f1/2})
by a constant times $N$ that is a matter of normalization.
We have determined the
coefficient $\mu$ only at $\beta=1$. In this case $\mu=-1/8$.
\item{\bf Special cases:}
\begin{itemize}
\item [-] A quasiharmonic
potential: $W=-|z|^2+V(z)+\overline{V(z)}$.
The density is uniform
inside the domain, $\sigma=1/\pi$, $F_{1/2}=F^{(1)}_1=0$,
\beq\label{FF}
F_1 =-\, \frac{1}{24\pi} \oint_{|w|=1}
(\phi \p_n \phi +2\phi )|dw|
\eeq
\item[-]  $\beta=2$. In this case
the ``classical" corrections ($F_{1/2}$ and
$F^{(1)}_{1}$) vanish for any potential $W$.
$F_{1}^{(0)}$ is given by (\ref{F10}).
\end{itemize}
\end{itemize}

These results, being combined with
the spectral determinant representation
(\ref{f1}), (\ref{K}),
suggest an explicit formula
for the determinant of the Laplace-Beltrami
operator in exterior domains.
For the interior domains, such a formula has been known
due to \cite{Polyakov,Alvarez}. It is called
the Polyakov-Alvarez formula. For simply-connected domains,
in can be written in terms of the conformal map of the
domain onto the unit disk. (A sketch of the
derivation is given in Appendix D.)
Our results suggest that for the exterior case
this formula remains basically the same, with
the interior conformal map being substituted by
the exterior one.
Below we list some formulas of this type.
Their rigorous proof is a challenging problem which is
beyond the scope of this paper.

\begin{itemize}
\item {\bf  The Polyakov-Alvarez formula for exterior
determinants (a conjecture)}:
\beq\label{PA}
\log \frac{\det (-e^{-2\chi}
\Delta_{\Dc})}{\det (-\Delta_{\Dc})}=
-\frac{1}{12\pi} \int_{\Dc} |\nabla \chi|^2 d^2z
+\frac{1}{6\pi} \oint_{\p \DD} \kappa \chi
ds -\frac{1}{4\pi}\int_{\Dc} \Delta \chi d^2 z
\eeq
(the divergent terms proportional to the area and
perimeter are omitted).
In the l.h.s.,
$\det (-\Delta_{\Dc})$ is the regularized spectral
determinant of the Laplace operator
in the Euclidean metric.
This determinant  is expressed entirely in terms of the boundary
value of the metric $e^\phi dwd\bar w=dzd\bar z$ induced
by the conformal map $w(z)$
of the exterior of $\DD$ onto the exterior of
the unit circle:
\beq\label{PA1}
\log \det (-\Delta_{\Dc})=
\frac{1}{12\pi} \oint_{\p\DD} \phi(\kappa+e^{-\phi}) ds
=\frac{1}{12\pi} \oint _{|w|=1}
(\phi \p_n \phi +2\phi )\, |dw|\, .
\eeq
Eqs. (\ref{PA}), (\ref{PA1}) give:
\begin{eqnarray}\label{PA2}
\log \det (-e^{-2\chi}
\Delta_{\Dc})&=& \displaystyle{
-\,\, \frac{1}{12\pi} \left [
\int _{|w|>1} |\nabla (\phi +\chi ) |^2 \, d^2 w
-2\oint _{|w|=1} \! (\phi +\chi ) \, |dw| \right ]}
\nonumber\\
&&-\,\, \frac{1}{4\pi}\int_{\Dc} \Delta \chi d^2z
\end{eqnarray}

\item{\bf Determinant formula (a conjecture).}
As a by-product we suggest another interesting formula for the
spectral determinant. Let
$t_k=\frac{1}{2\pi i k}\oint_{\p\DD}z^{-k}\bar z dz$
be harmonic moments of the exterior of the
domain $\DD$ and $\pi t_0$ be its area.
Let us assume that the domain
is such that all the moments at
$k>m$ vanish. Then, up to a constant, the
exterior spectral determinant is
$$
-12\,\log \det \left (- \Delta_{\Dc}\right ) =
\log\det_{m\times m} \left
(\frac{\p^3 F_0}{\p t_{0} \p t_{j} \p t_{k}}
\right )-(m^2 \! -\! 3m\! +\!
3) \, \frac{\p^2 F_0}{\p t_0^2}- (m\! -\! 1)\log
\bar t_m,
$$
where
$F_0$ is given by (\ref{free1})
with $\sigma=1/\pi$. Although the r.h.s. looks like a complex
number, it is actually real, as it will be clear
from the derivation below (see (\ref{det3}), (\ref{det4})).
\end{itemize}

\subsection{Comments and details}

\paragraph{The mean density in the leading order.}
In the leading order Eq. (\ref{L1}) reads
$$
2\beta \int \log |z-\zeta | \rho_0 (\zeta ) d^2
\zeta + W(z) =\Lambda\, ,
\quad \quad \Lambda=\frac{1}{2}(2-\beta )\hbar \lambda
$$
It must be valid in the domain where $\rho \neq 0$.
We assume that this domain (the support of the density)
is bounded, otherwise the normalization condition, i.e., that
the integral $\int \rho \, d^2 z$ be finite, can hardly be
satisfied.
We denote the support of the density by $\DD$.
In this paper, in order to avoid additional
technical complications,
we consider only the case of connected
domains $\DD$.

Upon taking the $z$-derivative of the equation above,
we get, for $z$ inside the domain $\DD$:
\beq\label{equilibr}
\p \vphcl (z) =\p W(z)
\quad \quad (z\in \DD )
\eeq
with
$$
\vphcl (z)= -\beta \int \log |z-\zeta |^2 \rhocl (\zeta )
d^2 \zeta
$$
being the 2D Coulomb potential created by the
equilibrium (``classical") configuration of the charges
characterized by the density $\rhocl$.
This equation just states that the total force
experienced by a charge at any point $z$, where
$\rhocl (z) \neq 0$,
is zero. Indeed, interaction with the other charges,
$\p \vphcl (z)$, is compensated by the
force $\p W(z)$ due to the external field.

The solution is conveniently expressed through
the function $\sigma$ (\ref{sigma}).
For the model to be well-defined, we assume that $\sigma (z) >0$
and tends to a positive constant as $|z|\to \infty$.
Applying $\p_{\bar z}$ to both sides of eq. (\ref{equilibr}),
one obtains the solution:
$\rhocl (z) = \sigma (z)/\beta$ inside $\DD$ and
$\rhocl (z) =0$ outside it.
In terms of the potential $\vphcl$ this solution reads
\beq\label{vphcl}
\vphcl (z) = -\int_{\DD} \log |z-\zeta |^2 \sigma (\zeta )
d^2 \zeta
\eeq
Assuming, without loss of generality, that $0 \in \DD$ and
$W(0)=0$, we fix
$\Lambda = -\vphcl (0)$, and so
$W(z)= \vphcl (z)- \vphcl (0)$.
Plugging this into (\ref{f0}), (\ref{f1/2}),
we find the leading contributions to the free energy
(\ref{free1}), (\ref{F1/2}).
$F_0$ is  the electrostatic energy
of the domain $\DD$ charged with the density $\sigma (z)$
and with a point-like compensating charge at the origin.
Different aspects
and applications of the functional (\ref{free1}) were
discussed in \cite{KKMWZ},
\cite{WZ2,Zles}, \cite{MWWZ}-\cite{Bertola}.

The most nontrivial part of the problem is to find
the shape of $\DD$.
It is determined by eq. (\ref{equilibr})
and by the normalization condition.
Using the Cauchy integral formula, we can write
these conditions in the form
\beq\label{shape}
\left \{
\begin{array}{l}
\displaystyle{
\oint_{\p \DD} \frac{\p W(\zeta )d\zeta}{z-\zeta}
=0}
\quad \quad \mbox{for all $z \in \DD$}
\\  \\
\displaystyle{
\int_{\DD}
\sigma (\zeta ) d^2 \zeta =\beta t_0 }
\end{array}
\right.
\eeq
To solve them for $\DD$ provided $W(z)$ and $t_0$
are given, amounts to a version of the inverse
potential problem in two dimensions. We assume that
the potential $W$ is such that the solution exists
and is unique. In general,
the solution is not available in an explicit form.
In this paper we do not address this question.
Our goal is to express corrections
to the free energy (\ref{free1}) in terms of the
domain $\DD$.

\paragraph{Pair correlation functions.}
The correlation functions in the leading order can be found
using the general variational formulas (\ref{var}),
where the exact free energy is replaced by $F_0$:
$$
\lim_{\hbar \to 0}
\lbracket \rho (z)\rbracket =
\frac{\delta F_0}{\delta W(z)} =\rhocl (z)\,,
\;\;\;\;\;
\lim_{\hbar \to 0}
\lbracket \rho (z_1)\rho(z_2)\rbracket _{c}=
\hbar^2 \, \frac{\delta \rhocl (z_1)}{\delta
W(z_2)}
$$
Basically, these are linear response relations used
in the Coulomb gas theory \cite{Jancovici}.
In this approximation,
the 2D Coulomb plasma is represented as a continuous charged
fluid, so the information about its discrete microscopic
structure is lost.
This is a good approximation
at distances much larger than the mean distance
between the charges.

Here are the results for the
correlation functions of the type
(\ref{tdens}). The mean value is obvious
from the result for the mean density given above:
\beq\label{1trace}
\hbar \beta \lbracket \mbox{tr} f \rbracket =
\int_{\DD}
\sigma (z) f(z) \, d^2 z \, + O(\hbar )
\eeq
The variation w.r.t. the potential yields
the connected parts of pair correlators \cite{WZ2}.
To present the result, we need
some elements of the Dirichlet boundary value problem.
Given a function $f(z)$,
let $f^H (z)$ be its
{\it harmonic continuation} from the boundary
of $\DD$ to its exterior, i.e., the solution
of the exterior Dirichlet boundary value problem.
The solution is given by the formula
\beq
\label{dirich}
f^H (z)=-\frac{1}{2\pi} \oint_{\p \DD}
f(\xi )\p_n G(z, \xi) |d\xi |
\eeq
where $G(z, \xi )$ is the Green function
of the domain $\Dc$.
It is the symmetric function
of two points uniquely determined
by the following properties:
$$
\Delta_z G(z, \zeta )=2\pi \delta (z-\zeta )
\quad \mbox{in $\Dc$}\,,
\;\;\;\;
G(z, \zeta )=0 \quad \mbox{if $z\in \p \DD$}
$$
For simply connected exterior domains $\Dc$
the Green function can be expressed through the conformal
map $w(z)$ from $\Dc$ onto the exterior
of the unit circle\footnote{Throughout the paper,
the map $w(z)$ is normalized as $w(z)=z/r + O(1)$ as
$z\to \infty$ with a real $r$.}:
\beq\label{Gw}
G(z , \zeta )=\log \left |
\frac{w(z )-w(\zeta )}{1-w(z ) \overline{w(\zeta )}}
\right |
\eeq
As $\zeta \to z$, it has the logarithmic singularity
$G(z, \zeta )\to \log |z-\zeta |$.
The connected pair correlator (\ref{tdens})
is then given by \cite{WZ2}:
\beq
\label{2trace}
4\pi \beta \lbracket \mbox{tr}f \, \mbox{tr} g\rbracket _{c}=
4\pi \beta \int f(z)\hat K^{-1}(z,z')\,g(z')d^2z d^2z'
=\int_{\DD} \nabla f \nabla g d^2 z-
\oint_{\p \DD} f \p_n g^H ds
\eeq
where $\nabla f$ is the gradient of the function $f$.
Alternatively, with the help of the
Green formula Eq. (\ref{2trace})
can be rewritten as a bulk integral (\ref{2trace1}),
where the integration goes
over the exterior domain and the entire plane:
\beq
4\pi \beta \int f(z)\hat K^{-1}(z,z')\,g(z')d^2z d^2z'
=-\int_{\Dc} \nabla (f-f^H)\nabla (g-g^H)   d^2 z +
\int_{\CC} \nabla f \nabla g d^2 z
\eeq
In particular, for the connected
correlation functions of the fields
$\varphi (z_1)$, $\varphi (z_2)$
this formula gives
(if $z_{1,2}\in \Dc$):
\beq
\label{2trace1}
\frac{1}{2\beta \hbar^2}\lbracket \varphi (z_1 )\varphi (z_2)
\rbracket_{c} =G(z_1 , z_2 ) -
G(z_1 , \infty )-G(\infty , z_2 )-
\log \frac{|z_1 - z_2 |}{r} +O(\hbar )
\eeq
where
$\log r= \lim_{z \to \infty}(
\log |z| +G(z , \infty ))$
is the (external) conformal radius of $\DD$.

\paragraph{Spectral determinants.}
The relation (\ref{K}) between the spectral determinants,
\beq
\log \det \hat K=\log \det (-e^{-2\chi}\Delta_{\Dc})-
\log \det (-e^{-2\chi}\Delta_{\CC})\,,
\eeq
can be understood using formula (\ref{2trace}).
The presence of the bulk and boundary
terms in the r.h.s.
suggests to separate boundary and bulk values
of the functions $f$
and $g$. Specifically, we write $f = f_H + \tilde f$, where
$f_H$ is the harmonic continuation of the $f$ to the interior
of $\DD$ while the boundary value of $\tilde f$ is zero,
and similarly for $g$.
Using the Green theorem, one can see that the
bulk and boundary contributions
completely separate\footnote{We deliberately keep the
factor $e^{2\chi}=\pi\sigma$
in  some formulas  in order to
emphasize the fact that the Laplace-Beltrami
operator is an invariant
operator with respect to the metric
$\pi \sigma$ written in the Weyl gauge.
Although the metric cancels in the formula below,
it appears in the spectral determinants.}:
$$
4\pi \beta \int f\, \hat K^{-1}  g
= \,\int_{\DD} \tilde f \left (-e^{-2\chi}
\Delta \right )\tilde g\, e^{2\chi} d^2 z +
\oint_{\p \DD} f \hat N g \, ds
$$
Here
$$
\hat N g(z)=\p_{n}^{+}g_H (z) - \p_{n}^{-}g^H (z)
$$
is the Neumann jump operator ($\p_{n}^{\pm}$ are normal derivatives
from inside and outside respectively) which sends a function
on the boundary to the difference of normal derivatives of its
harmonic continuations inside and outside the domain $\DD$.
This means that the operator $\hat K^{-1}$ is the direct sum
of the Laplace operator
in $\DD$ (with the Dirichlet b.c.)
and the Neumann jump operator. Therefore,
$\det (\hat K^{-1})\simeq
\det (-e^{-2\chi}\Delta_{\DD}) \, \det \hat N$, and so
\beq\label{zqdetdet}
F^{(0)}_1=\frac{1}{2}\log
\det (-e^{2\chi}\Delta_{\DD})+ \,\frac{1}{2}\log \det \hat N
\eeq
(we omit an irrelevant factor).
This expression can be simplified by means of the following
relation between the
properly regularized functional determinants \cite{Zelditch}:
$$
\log \det (-e^{-2\chi}\Delta_{\DD}) +
\log  \det \hat N +
\log \det (-e^{-2\chi}\Delta_{\Dc})=
\log P(\DD ) +\log \det (-e^{-2\chi}\Delta_{\CC})
$$
Here
$P(\DD )
=\oint_{\p \DD} \sqrt{\pi \sigma}\, ds$
is the perimeter of $\DD$ in the metric
$\pi \sigma$.

This relation, rigorously
proven in \cite{Zelditch},
is known in the mathematical literature as
the ``surgery formula". It is clearly motivated
by the ``cut and paste" physical arguments.
Consider the free bosonic theory in the whole
plane with the quadratic action
$S_0 =\int |\nabla X|^2 d^2 z$ (for brevity, we
do not indicate the metric explicitly).
The path integral $\int [DX]\exp (-S_0[X])$ is
equal to $(\det (-\Delta_{\CC}))^{-1/2}$.
On the other hand, let us fix a domain $\DD$
and represent the action as
$$
S_0 =
\int_{\DD} |\nabla X|^2 d^2 z
+\int_{\Dc} |\nabla X|^2 d^2 z
$$
Decompose the field $X$ inside $\DD$ into the
sum of the field $\tilde X$ such that $\tilde X =0$
on $\p \DD$
and the harmonic field $X_H$: $X=\tilde X + X_H$.
Let $X=\tilde X + X^H$ be the similar decomposition
for the field $X$ outside $\DD$ (we then have
$X_H = X^H$ on $\p \DD$). It is easy to see that
these fields separate in the action as follows:
$$
S_0 [X] =-\int_{\DD} \tilde X \Delta \tilde X \, d^2 z
-\int_{\Dc} \tilde X \Delta \tilde X \, d^2 z
+\oint_{\p \DD} X_H \hat N X_H \, ds
$$
This separation implies the surgery formula.
The term $\log P(\DD )$ is due to the zero mode
of the Neumann jump operator which we did not take
into account.

Hence, $F^{(0)}_1$, the ``quantum" part
of $F_1$, is to be identified with
$\frac{1}{2}\log \det (-e^{-2\chi}
\Delta_{\CC}) -
\frac{1}{2}\log
\det (-e^{-2\chi}\Delta_{\Dc})$,
where the determinants have to be properly
regularized.
Some additional efforts are required to refine these
arguments. In particular, one should justify the
choice of the background metric in $\DD$ and
take care of the zero mode of the Neumann jump operator.
This is beyond the scope of the present paper.
The next section provides an alternative derivation
of $F_1$.

\section{Corrections to the free energy from the  loop equation}

In the previous sections we have found that
the asymptotic expansion
of the partition function as $\hbar \to 0$
has the form
\beq\label{hbarexp1}
Z_N = N! \hbar^{\frac{1}{2}(2-\beta )N} e^{\gamma N}
\exp \left ( \frac{F_0}{\hbar^2} +
\frac{F_{1/2}}{\hbar} +F_1 +\sum_{k\geq 3} \hbar^{k-2}F_{k/2}
\right )
\eeq
The corresponding expansions of the mean values of
$\rho$ and $\varphi$ are
\beq\label{meanrho}
\lbracket \rho (z)\rbracket =
\rhocl (z) +\hbar \rho_{1/2} (z) +\hbar^2 \rho_1 (z)
+O(\hbar^3)
\eeq
\beq\label{meanphi}
\lbracket \varphi (z)\rbracket =
\vphcl (z) +\hbar \varphi_{1/2} (z) +\hbar^2 \varphi_1 (z)
+O(\hbar^3)
\eeq
where $\rho_{i}(z)=\delta F_i /\delta W(z)$
and $\varphi_{i}(z)=-\beta \int \rho_i (\zeta )
\log |z-\zeta |^2 d^2 \zeta$.
The terms $F_0$ and $F_{1/2}$ are given by (\ref{free1})
and (\ref{F1/2}) respectively.
The higher corrections are due to fluctuations of
the charged particles around the equilibrium state.
In principle,
they can be found by expanding the loop
equation (\ref{loopeq})
\beq\label{loopeqa}
\frac{1}{2\pi}\int \frac{\p W(\zeta )
\lbracket \Delta \varphi (\zeta )\rbracket}{z-\zeta}
\, d^2 \zeta =
\lbracket (\p \varphi (z))^2 \rbracket
+(2\! -\! \beta )\hbar \, \lbracket \p^2 \varphi (z)\rbracket
\eeq
in powers of $\hbar$ and  solving the
inhomogeneous linear integral equations
obtained in this way. A
similar approach has been developed in  the case of  Hermitian
matrix ensembles \cite{herm1}. This is what we are going to do
in this section. We restrict ourselves by
$F_{1/2}$ and $F_1$.
Calculations of higher order corrections are   rather tedious.

\subsection{Iteration  of the loop equation}

As it was already pointed out,
the main contribution to the partition function
as $\hbar \to 0$ comes from a configuration,
where the charges are ``frozen" at their equilibrium
positions.
Correspondingly, the averages
take their ``classical" values
$\lbracket \varphi (z) \rbracket = \vphcl (z)$,
and multipoint correlators factorize in the
leading order:
$\lbracket \p \varphi (z) \,
\p \varphi (z')\rbracket =
\p \vphcl (z) \p\vphcl (z')$.
Under this assumption, the loop equation
becomes a closed relation for $\vphcl$:
\beq\label{loopeq2}
\frac{1}{2\pi} \int \frac{\p W( \zeta )\Delta
\vphcl (\zeta )}{z-\zeta }\, d^2 \zeta \,
= \, \Bigl ( \p \vphcl (z)\Bigr )^2
\eeq
where we have omitted the last term in
(\ref{loopeqa}) which is
of the next order in $\hbar$.
Let us apply $\p_{\bar z}$ to both sides of the equation.
This yields: $\p W(z) \Delta \vphcl (z)=
\p \vphcl (z) \Delta \vphcl (z)$.
Since $\Delta \vphcl (z) \propto \rhocl (z)$
we obtain
\beq\label{large1}
\rhocl (z) \, \left [ \p \vphcl (z) -\p W(z)\right ]=0
\eeq
This equation should be solved with the
additional constraints
$\int \rhocl (z) d^2 z =t_0$ (normalization) and
$\rhocl (z)\geq 0$ (positivity).
The equation tells us that
either $\p \vphcl (z)=\p W(z)$ or
$\rhocl (z)=0$.
Applying $\p_{\bar z}$,
we get $\Delta \vphcl (z)=\Delta W(z)$.
This gives the solution for $\rhocl$ and
$\vphcl$ already obtained in Sec. 2 by less
formal arguments (see (\ref{vphcl})).

Now we are in a position to develop the $\hbar$-expansion
of the loop equation (\ref{loopeq}).
First of all, we rewrite it identically in the form
$$
\frac{1}{2\pi}\int L(z, \zeta )
\lbracket \Delta \varphi (\zeta )\rbracket d^2 \zeta =
(\p \vphcl (z))^2 -
\left (\p  \left ( \lbracket \varphi (z) \rbracket
\phantom{^2} \!\! - \, \vphcl (z) \right ) \right )^2
$$
$$
- \lbracket \left ( \, \p \,\, [ \varphi (z) \phantom{^2} \! -\,
\lbracket  \varphi (z)  \rbracket \, ] \, \right )^2
\rbracket -
(2\! -\! \beta )\hbar  \lbracket \p^2 \varphi (z) \rbracket
$$
which is ready for the $\hbar$-expansion.
Here
\beq\label{loopkern}
L(z, \zeta ) =\frac{ \p W(\zeta ) - \p \vphcl (z)}{\zeta -z}
\eeq
is the kernel of the integral operator in the l.h.s. (the
``loop operator"). The zeroth order in $\hbar$ gives
equation (\ref{large1}) which implies the familiar result
$\vphcl (z)=-\int_{\DD} \log |z-\zeta |^2
\sigma (\zeta )d^2 \zeta$ for the $\vphcl$.
To proceed, one should insert the series (\ref{meanphi})
into the loop equation and separate
terms of order $\hbar$, $\hbar^2$ etc. The terms of order
$\hbar$ and $\hbar^2$ give:
\beq\label{exploop}
\begin{array}{l}
\displaystyle{
\frac{1}{2\pi}\int L(z, \zeta )
 \Delta \varphi_{1/2} (\zeta ) d^2 \zeta =
-(2 \! -\! \beta ) \p^2 \vphcl (z)}
\\ \\
\displaystyle{
\frac{1}{2\pi}\int L(z, \zeta )
 \Delta \varphi_1 (\zeta ) d^2 \zeta =
-\left [ \left ( \p \varphi_{1/2}(z)\right )^2 +
(2 \! -\! \beta ) \p^2 \varphi_{1/2} (z)\right ]
-\omega (z)}
\end{array}
\eeq
where
\beq\label{omega1}
\omega (z)= \lim_{\hbar \to 0} \left [ \hbar^{-2}
\lim_{z'\to z} \lbracket
\p \varphi (z) \, \p \varphi (z') \rbracket_c \right ]
\eeq
is the connected part of the pair correlator at merging
points.

The expansion of the loop equation can be continued
order by order. In principle, this gives
a recurrence procedure to determine the coefficients
$\varphi_k (z)$.
However, each step requires solving integral equations
in the plane, that is not easy
to do explicitly.
Another difficulty is that
in general one can not extend these equations
to the interior of the support of the density
because the $\hbar$-expansion may break down
or change its form there.
Indeed, in the domain where the density is macroscopically
nonzero,
the microscopic structure of the gas becomes essential, and one
needs to know correlation functions at small scales.
Nevertheless, at least in the first two orders in $\hbar^2$
the equations above can be solved  assuming that
$z \in \Dc$. Note that in this region all
the functions $\varphi_k (z)$ are harmonic.
If these functions are known, the corresponding
expansion coefficients of the free energy in
can be obtained
by ``integration" of the
variational formulas (\ref{var}).

\subsection{The solution for $F_{1/2}$}

We start with the order $\hbar$. We need to solve
the first equation in (\ref{exploop}). Using (\ref{rhophi}),
(\ref{var}), we rewrite it in the form
\beq\label{S0}
\int \frac{\delta F_{1/2}}{\delta W(z )}\,
L(a, z )\, d^2 z =
\frac{2-\beta}{2\beta} \, \int_{\DD}
\frac{\sigma (z )d^2 z}{(a-z )^2}
\eeq
where $a$ is an arbitrary point in $\Dc$.
Using the variational technique developed in
\cite{WZ2} (see also Appendix A),
one can verify that $F_{1/2}$ given in
eq. (\ref{F1/2}) does solve this equation.

\paragraph{$F_{1/2}$ from the loop equation.}
Here we give some details of this calculation.
Exactly the same scheme is used in the next
subsection while solving the loop equation for
$F_1$.

It is convenient to use the notation (\ref{sigma})
$\chi (z) =\log \sqrt{\pi \sigma (z)}$
in terms of which $F_{1/2}= - \frac{2-\beta}{\pi \beta}
\int_{\DD} e^{2\chi}\chi \, d^2 z$. The variation of this
functional reads
\beq\label{varF12}
\delta F_{1/2} =-\frac{2-\beta}{\pi \beta}
\left [ \oint_{\p \DD} e^{2\chi}\chi \, \delta n \, ds
+ \int_{\DD} e^{2\chi} (2\chi +1)\delta \chi \, d^2 z
\right ]
\eeq
Here
\beq\label{deltan}
\delta n(z)= \frac{\p_n (\delta W (z) \! -\!
\delta W^H (z))}{4\pi \sigma (z)}
\eeq
is the normal displacement of the boundary (Fig.~1) under
variation of the potential, with the
convention that $\delta n >0$ for
outward displacement (for the proof see \cite{WZ2,Zles})
and
\beq\label{chivar}
\delta \chi (z)= - \, \frac{\Delta \delta W(z)}{8\pi \sigma (z)}
\eeq
Note that the l.h.s. of (\ref{S0}) has the meaning of the
variation of $F_{1/2}$ under a small change of the potential
proportional to $L(a,z)$, $\delta W(z)\propto L(a,z)$
($a$ plays the role of a parameter). Therefore,
the result is given by (\ref{varF12})
where $\delta n$ and
$\delta \chi$ are taken from (\ref{deltan}), (\ref{chivar})
with $\delta W(z)=L(a,z)$.
It remains to plug the explicit form of the $L(a,z)$
(\ref{loopkern}) and simplify the result.
In the course of this calculation, the frequently used
formulas are:
\beq\label{g6}
\Delta_z L (a,z)=4\pi \p_z \left ( \frac{\sigma (z)}{a-z} \right )
\eeq
and
\beq\label{g7}
\oint_{\p \DD} f(z) \p_n (L (a,z)\! -\! L^{H}(a,z)) \, ds =
2\pi i \oint_{\p \DD}  \frac{f(z)\sigma (z)}{a-z} \, d\bar z
\eeq
(for any smooth function $f$).
It is implied that $z\in \DD$, $a\in \Dc$.
In these formulas,
the Laplace operator and the harmonic continuation
are applied to $z$.
We have:
$$
\int \frac{\delta F_{1/2}}{\delta W(z )}\,
L(a, z )\, d^2 z
$$
$$
= \, -\, \frac{2-\beta}{4\pi \beta}
\left [
\oint_{\p \DD} \chi (z) \p_n (L(a,z)\! -\! L^H (a,z))ds -
\frac{1}{2} \int_{\DD} (2\chi (z) +1) \Delta L(a,z) d^2 z \right ]
$$
$$
=\, -\, \frac{2-\beta}{4\pi \beta}
\left [-i \oint_{\p \DD} \frac{\chi (z)\sigma (z)}{z-a}d\bar z
+ \int_{\DD}(2\chi (z) +1)\p_z \left ( \frac{\sigma (z)}{z-a}\right )
d^2 z \right ]
$$
After transforming the first (contour) integral
to the integral over the domain $\DD$,
$$
\oint_{\p \DD} \frac{\chi (z)\sigma (z)}{z-a}d\bar z
=-2i \int_{\DD}\p_z \left ( \frac{\chi (z)\sigma (z)}{z-a}\right )
d^2 z
$$
one can see that
the result is indeed equal to the r.h.s. of (\ref{S0}).

\begin{figure}[tp]
\epsfysize=4cm
\centerline{\epsfbox{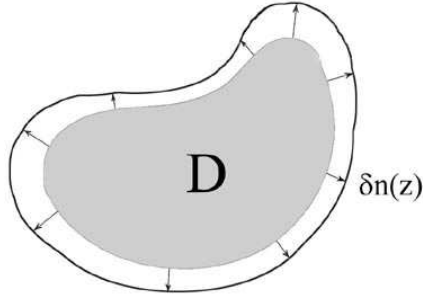}}
\caption{\sl The normal displacement of the boundary.}
\label{fi:deltan}
\end{figure}

\paragraph{The results for $\rho_{1/2}$ and $\varphi_{1/2}$.}
The corrections to the mean values
of $\rho$ and $\varphi$ are
conveniently expressed through the function $\chi$
(\ref{sigma})
and its harmonic continuation $\chi^H (z)$
from the boundary of
$\DD$ to its exterior.
In terms of these functions
\beq\label{S2}
\rho_{1/2}(z)=
\frac{\delta F_{1/2}}{\delta W(z)}=
\frac{2-\beta}{4\pi \beta} \left (
\Theta (z; \DD )\Delta \chi (z) -
\delta  (z; \p \DD )\,
\p_n (\chi (z)\! -\! \chi^H (z))-
\frac{1}{2} \delta ' (z; \p \DD )
\right )
\eeq
Here $\Theta (z; \DD)$ is the characteristic function
of the domain $\DD$ (1 if $z\in \DD$ and 0 otherwise),
$\delta (z;\p \DD)$ is
the $\delta$-function with the support
on the boundary ($\int f(z)\delta (z; \p \DD)d^2z
 =\oint_{\p \DD} f(z)|dz|$
for any smooth function $f$) and
$\delta '(z; \p \DD )$ is
its ``normal derivative'', i.e. a function such that
$\int f(z)\delta '(z; \p \DD )
d^2z =-\oint_{\p \DD}\p_n f(z)|dz|$.
The singular function $\rho_{1/2}$ is to be understood as
being integrated with any smooth test function.
The first term in the r.h.s. describes the change
of density in the bulk.
The second one
(the ``simple layer") describes
a shift of the boundary of the domain $\DD$.
The last one, the ``double layer",
means smoothing off the edge of the density support
(for $\beta \neq 2$).
Using the operator $\hat K$ introduced in Sec. 3,
the correction $\rho_{1/2}$ can be written as
$$
\rho_{1/2}(z)=-\, (2 \! -\! \beta ) [\hat K^{-1}\chi ](z)
-\frac{2\! -\! \beta}{8\pi \beta}\, \delta '
(z;\p \DD )
$$
At $\beta =2$ the correction
$\rho_{1/2}$ vanishes.
The first correction to $\lbracket \varphi (z) \rbracket$
reads:
\beq\label{S3}
\varphi_{1/2}(z)=\left \{
\begin{array}{ll}
-(2\! -\! \beta )\left [ \chi (z) -\chi^H (\infty ) +\frac{1}{2}
\right ]\,, & \;\; z\in \DD
\\ & \\
-(2\! -\! \beta )\left [ \chi^H (z) -\chi^H (\infty )
\right ]\,, & \;\; z\in \Dc
\end{array} \right.
\eeq
Due to the double layer, this function is discontinuous across the
boundary.
Therefore, for $z\in \Dc$ we have: $\p \varphi_{1/2}(z)=-(2\! -\!
\beta ) \p \chi^H (z)$.

\subsection{The solution for $F_1$}

Now we have to solve the second equation in
(\ref{exploop}), where $\p \varphi_{1/2}=-(2\! -\!
\beta ) \p \chi^H$
and $\omega (z)$ is still to be
found.
If the point $z$ is in $\Dc$,
then eq. (\ref{2trace1}) yields \cite{WZ2}:
$$
\lbracket
\p \varphi (z) \, \p \varphi (z') \rbracket_c
=2\beta \hbar^2 \p_{z}\p_{z'} \Bigl (
G(z, z') -\log |z-z'|\Bigr ) +O(\hbar^3 ))
$$
Since the r.h.s. is regular for all $z,z' \in \Dc$,
the points can be merged without any difficulty and the
result does not depend on the particular limit $z' \to z$.
Using the expression of the Green function (\ref{Gw}) through the
conformal map $w(z)$,
we obtain
$$
12\lim_{z' \to z}\p_z \p_{z'} \Bigl (
G(z, z') -\log |z-z'| \Bigr ) =
\frac{w'''(z)}{w'(z)}- \frac{3}{2}
\left (\frac{w''(z)}{w'(z)}\right )^2
=\{w; \, z\}
$$
where we use the standard notation for the
Schwarzian derivative.
The function $\omega$ is thus given by
\beq\label{omega}
\omega (z)
=\frac{\beta}{6}\{w; \, z\}
\eeq
After plugging the above result for $\varphi_{1/2}$,
the second equation of (\ref{exploop}) acquires the form:
\beq\label{S4}
\int \frac{\delta F_{1}}{\delta W(\zeta )}\,
L(z, \zeta )\, d^2 \zeta =
\alpha^2 \left [
\bigl (\p \chi^H (z)\bigr )^2 - \p^2 \chi^H (z) \right ]
+\frac{1}{12}\{ w ; \, z\}
\eeq
where
$
\alpha =\sqrt{\frac{2}{\beta}}-
\sqrt{\frac{\beta}{2}}$ (as is in (\ref{sigma})).
We consider this equation for $z\in \Dc$, where
the both sides are harmonic functions.

Our strategy is as follows.  We make a guess
for $F_1$ and then show that
the variational derivative obeys (\ref{S4}). More precisely,
let $I^{{\rm trial}}$ be a trial functional of $\DD$
(and thus of $W$) represented as an
integral over the domain $\DD$ or its boundary.
The form of the trial functionals
is suggested by the path integral
arguments of the previous section.
Since $F_1$ has dimension $0$, we
consider only dimensionless functionals.
We want to find
$\int \delta I^{{\rm trial}}/\delta W(z) L(a,z ) \,d^2 z$
and compare with the r.h.s. of (\ref{S4})
(again, $a$ is a
point in $\Dc$). The latter quantity can be computed
by the method outlined in the previous subsection.

A comment on the meaning
of the calculations below is in order.
They consist of recognizing that the
l.h.s. of eq. (\ref{S4}) is a variation of $F_1$ over
the holomorphic
component of the metric and then applying the Polyakov-Alvarez
formula (\ref{PA}). We plan to elaborate on this point
elsewhere.

The structure of eq. (\ref{S4})
suggests to find different terms of the
solution separately.
Let us start with the term proportional to $\alpha^2$.
Here are the main steps of the calculations.
Consider the functional
\beq\label{S5}
I^{(1)}=\frac{1}{4\pi}\left (
\int_{\DD} |\nabla \chi |^2 d^2 z -\oint_{\p \DD}
\chi \p_n \chi^H ds \right )
\eeq
Its variation is (see Appendix B):
\beq\label{S5a}
\delta I^{(1)}=\frac{1}{4\pi}
\oint_{\p \DD} \left [ \p_n (\chi -\chi^H )\right ]^2
\delta n \, ds +\frac{1}{2\pi}\oint_{\p \DD}
\delta \chi \p_n (\chi -\chi^H )\, ds
-\frac{1}{2\pi}\int_{\DD} \delta \chi \Delta \chi \, d^2 z
\eeq
In the same way as for $F_{1/2}$ one can check that
$$
\int \frac{\delta I^{(1)}}{\delta W(z )}\,
L(a, z )\, d^2 z =
\bigl (\p \chi^H (a)\bigr )^2 - \p^2 \chi^H (a)
$$
so $F_1$ is expected to contain the term
$\alpha^2 I^{(1)}$ (compare
with $F^{(1)}_1$ given by (\ref{F111})).
This contribution is of the ``classical" nature
since no fluctuations of the charges positions are taken
into account.

Next, we consider
\beq\label{S6}
I^{(2)}=-\, \frac{1}{2\pi}\oint_{|w|=1}(\phi \p_n \phi +2\phi )|dw|
\eeq
 (we remind that
$\label{phi}
\phi (w)=\log |z'(w)|
$
where $z(w)$ is the conformal map from the
exterior of the unit circle onto $\Dc$ inverse to
the $w(z)$).
The variation of this functional is found in Appendix B
(eq. (\ref{varI1})):
\beq\label{varI1a}
\delta I^{(2)}=\frac{1}{2\pi}
\oint  \left (\nu^2 (z) \{w;z\} +\overline{\nu^2(z) \{w;z\}}
-2\kappa^2(z) \right ) \delta n(z) \,ds
\eeq
Here
\beq\label{nuw}
\nu (z)= |w'(z)|\frac{w(z)}{w'(z)}
\eeq
is the normal unit vector
to the boundary\footnote{It is worthwhile to mention
here the useful formulas for normal and tangential
derivatives:
$\p_n f =\nu  \p_z f +\overline{\nu} \p_{\bar z}f$,
$\p_s f =i\nu  \p_z f -i\overline{\nu} \p_{\bar z}f$ and
for the line element on the boundary curve:
$ds =\frac{dz}{i\nu (z)}=i\nu (z) d\bar z$.} and
$$
\kappa (z) =\p_n \log \left |\frac{w(z)}{w'(z)}\right |
$$
is the local curvature of the boundary (counted w.r.t.
the outward pointing normal vector). Using the rules explained
above, we get:
$$
\int \frac{\delta I^{(2)}}{\delta W(z )}\,
L(a, z )\, d^2 z =-\,
\frac{1}{4\pi i}
\oint_{\p \DD}
\frac{\nu^2  \{w;z\} +\overline{\nu^2 \{w;z\}}
-2\kappa^2}{a-z}\, d\bar z
$$
Now consider the functional
\beq\label{S6a}
I^{(3)}= \frac{1}{2\pi}
\int_{\DD} |\nabla \chi |^2 d^2 z +
\frac{1}{\pi}\oint_{\p \DD}
\kappa \chi ds
\eeq
with the variation
$$
\delta I^{(3)}=\frac{1}{2\pi}\oint_{\p \DD}
\left ( |\nabla \chi |^2 \! -\! 2\p_{s}^{2}\chi
\! +\! 2\kappa \p_n \chi \right )
\delta n \, ds
\! +\!  \, \frac{1}{\pi}
\oint_{\p \DD} (\kappa \! + \! \p_n \chi )\delta \chi \, ds
-\frac{1}{\pi} \int_{\DD} \Delta \chi \, \delta \chi \, d^2 z
$$
The variational derivative $\delta I^{(3)}/\delta W$ looks
rather complicated. However, for its convolution with
$L(a,z)$ one obtains a surprisingly simple result
$$
\int \frac{\delta I^{(3)}}{\delta W(z )}\,
L(a, z )\, d^2 z =
-\, \frac{1}{2\pi} \oint_{\p \DD} \frac{\kappa ds}{(a-z)^2} \, =\,
\frac{1}{2\pi i} \oint_{\p \DD}
\frac{\kappa^2 +i\kappa '}{a-z}\, d\bar z
$$
(note that $\chi$ cancels!). Combining it with the corresponding
result for $I^{(2)}$, we obtain:
$$
\int \frac{\delta (I^{(2)} \! -\! I^{(3)})}{\delta W(z )}\,
L(a, z )\, d^2 z =
-\, \frac{1}{4\pi i}
\oint_{\p \DD}
\frac{\nu^2  \{w;z\} +\overline{\nu^2 \{w;z\}}
+2i \kappa '}{a-z}\, d\bar z
$$
$$
=\, - \, \frac{1}{2\pi i}
\oint_{\p \DD} \frac{\nu^2 (z)  \{w;z\}}{a-z}\, d\bar z \, = \,
\frac{1}{2\pi i} \oint_{\p \DD}
\frac{\{w;z\}}{a-z}\, d z \,\, = \,\,
\{w(a); \, a\}
$$
where (\ref{curv4}) has been used.

At last, the functional
\beq\label{S8}
I^{(0)}=\frac{1}{2\pi}\int_{\DD} \Delta \chi \, d^2 z
\eeq
with the variation
$$
\delta I^{(0)}=\frac{1}{2\pi}\oint_{\p \DD}
\left ( \Delta \chi \, \delta n + \p_n (\delta \chi )\right ) ds
$$
is a ``zero mode'' of the loop operator:
$$
\int \frac{\delta I^{(0)}}{\delta W(z )}\,
L(a, z )\, d^2 z =0
$$

Summing all the contributions
with appropriate coefficients that follow from the
r.h.s. of the loop equation,
we thus find $F_1$:
$$
F_1 =F^{(1)}_1+F^{(0)}_1=
\alpha^2 I^{(1)} +
\frac{1}{12}(I^{(2)} \! -\! I^{(3)}) +\mu I^{(0)}
$$
where the coefficient $\mu$ can not be determined
from the loop equation restricted to $\Dc$. (In the
whole plane, the loop operator does not have zero modes
but we have to be restricted to $\Dc$ because of unknown
properties of correlation functions at small distances
in the bulk.) For $\beta =1$ the coefficient $\mu$ can be fixed
by comparison with the explicit solution for centrosymmetric
potentials (see Appendix C):
$$
\mu = -\frac{1}{8}
\quad \quad (\mbox{at $\beta =1$})
$$

With the help of the Green theorem, we present the
result in the form appearing in Sec.\ref{Su}:
\beq\label{ff3}
\begin{array}{lll}
F_1 &=& \displaystyle{
\frac{1}{24\pi} \left [
\int _{|w|>1} |\nabla (\phi +\chi ) |^2 \, d^2 w
-2\oint _{|w|=1} \! (\phi +\chi ) \, |dw| \right ]}
\\ &&\\
&&\displaystyle{
+\,\,\, \frac{\alpha^2}{4\pi}\left [ \int _{\DD} |\nabla \chi |^2
d^2 z -\oint _{\p \DD}\chi \p_n \chi^H \, ds\right ] \, }
\\ &&\\
&&\displaystyle{
+\,\,\,\frac{\mu}{2\pi} \int _{\DD} \Delta \chi d^2 z
-\frac{1}{24\pi}\int_{\CC} |\nabla \chi |^2 \, d^2 z}.
\end{array}
\eeq
where $\chi$ in the first two terms
is regarded as a function of $w$ via
$\chi =\chi (z(w))$.
Let us also list some equivalent forms:
\beq\label{ff1a}
\begin{array}{lll}
F_1 &=& \displaystyle{
-\,\, \frac{1}{24\pi} \oint _{|w|=1}
(\phi \p_n \phi +2\phi )\, |dw|} \,
\\ &&\\
&&\displaystyle{
-\,\,\,\, \frac{1-6\alpha^2}{24 \pi}
\left[ \int _{\DD} |\nabla \chi |^2
d^2 z +2\oint _{\p \DD}\kappa \chi \, ds\right ] \, }
\\ &&\\
&&\displaystyle{
-\,\,\,\, \frac{\alpha^2}{4\pi}\oint _{\p \DD} \chi
(\p_n \chi^H +2\kappa )\, ds \, + \,\frac{\mu}{2\pi}
\oint_{\p \DD}\p_n \chi \, ds ,}
\end{array}
\eeq
\beq\label{ff4}
\begin{array}{lll}
F_1 &=&\displaystyle{\frac{1-6\alpha^2}{24\pi}
\left [ \int_{|w|>1} |\nabla (\phi +\chi )|^2 d^2w
-2\oint_{|w|=1} (\phi +\chi )|dw| -\int_{\CC}|\nabla
(\phi +\chi )|^2 d^2z\right ] }
\\ &&\\
&&\displaystyle{+
\,\,\,\,
\frac{\alpha^2}{4\pi}
\left [ \int_{|w|>1} |\nabla (\phi +\chi^H )|^2 d^2w
-2\oint_{|w|=1} (\phi +\chi^H )|dw| \right ]
+\frac{\mu}{2\pi}
\int_{\DD}\Delta \chi d^2z \, .}
\end{array}
\eeq

\section{Models with quasiharmonic potentials}

The models with $W$ of the form
$$
W=-|z|^2 +V(z)+\overline{V(z)}
$$
(``quasiharmonic potentials") generalize the Ginibre-Girko
ensemble \cite{GG}.
Note that for potentials
of this form the integral (\ref{I1}) diverges unless
$V$ is quadratic or logarithmic with suitable coefficients.
The simplest way to give sense to the integral
when it diverges at infinity is
to introduce a cut-off, i.e., integrate over
a suitably chosen big but finite domain in the plane.
Then the large $N$ expansion is well-defined and the
results for the general potential presented above
still make sense.
For details and rigorous proofs
see \cite{rigorous}.

In the case of quasiharmonic potentials
the formula for $F_1$ drastically simplifies since
the function $\chi$ vanishes, and so
only the first integral in (\ref{ff1a}) survives:
\beq\label{ff6}
F_1 =
-\,\, \frac{1}{24\pi} \oint _{|w|=1}
(\phi \p_n \phi +2\phi )\, |dw|
\eeq
In the particular case $\beta =1$,
$W(z)=-z\bar z$ the formula yields
$F_1 =-\frac{1}{12}\log t_0$ that coincides with the
result of \cite{FGIL} obtained by a direct calculation.

According to the conjecture
of Section 4.1, eq. (\ref{ff6}) can be understood as
the formula for the regularized
determinant of the Laplace operator
$\Delta_{\Dc} =4\p_z \p_{\bar z}$ in
the exterior domain $\Dc$
with the Dirichlet boundary conditions:
\beq\label{f3}
F_1 =-\, \frac{1}{2} \log \det \left (- \Delta_{\Dc}\right )
\eeq
The first term is the bulk contribution (for the metric
induced by the conformal map it reduces to a boundary integral),
while the second term is a net boundary term.
The ``classical" contribution $F^{(1)}_1$ to $F_1$
vanishes in this case.

\subsection{The case of rational $\p V(z)$ (quadrature
domains)}

There is a special class of domains
for which our result (\ref{ff6}) can be made more
explicit.
Consider domains such that $z'(w)$ is a rational function,
$$
z'(w)=r\prod_{i=0}^{m-1}\frac{w-a_i}{w-b_i}
$$
In the mathematical literature, they
are called quadrature domains \cite{quadrature}.
One can show that quadrature domains
are density supports for the models with
potentials such that $\p V(z)$ is a rational function.
All the points $a_i$ and $b_i$
must be inside the unit circle, otherwise the map
$z(w)$ is not conformal. As $w\to \infty$,
$z(w)$ can be represented as a Laurent series of the form
$z(w)= rw + u_0 +O(w^{-1})$.
On the unit circle we have
$|dw| =\frac{dw}{iw}$
and $\phi (w) =\frac{1}{2}( \log z'(w) +
\log \bar z'(w^{-1}))$,
where the first and the second term
(the Schwarz reflection) are analytic
outside and inside it, respectively.
(Recall that $\bar z (w) \equiv
\overline{z(\bar w)}$ and
our notation $\bar z'(w^{-1})$ means
$d \bar z(u)/du$ at the point $u=w^{-1}$.)
Plugging this into (\ref{ff6}), we get:
$$
\begin{array}{ll}
F_1 \, =&\displaystyle{-\, \frac{1}{24\pi i}\oint_{|w|=1}
\log z'(w)
\left [ \frac{1}{2}\p_w \log z'(w)+\frac{1}{w}\right ] dw}\, -
\\&\\
&\displaystyle{- \, \frac{1}{24\pi i}\oint_{|w|=1}
\log \bar z'(w^{-1}) \frac{dw}{w} -
\frac{1}{48\pi i}\oint_{|w|=1}
\log \bar z'(w^{-1}) \frac{z''(w)}{z'(w)}dw}
\end{array}
$$
The integrals can be calculated by taking residues
either outside or inside the unit circle.
The poles are at $\infty$, at $0$, and at the points
$a_i$ and $b_i$.
The result is
\beq\label{PA11}
F_1 =-\frac{1}{24}\left ( \log r^4 +
\sum_{z'(a_i)=0}\log \bar z'(a_{i}^{-1})
-\sum_{z'(b_i)=\infty}\log \bar z'(b_{i}^{-1})
\right )
\eeq
If the potential $V(z)$ is polynomial,
$V(z)=\sum_{k=1}^mt_k z^k$,
i.e., $t_k =0$ as $k > m$ for some $m>0$, then
the series for the conformal map $z(w)$ truncates:
$z(w)= rw +\sum_{l=0}^{m-1}u_l w^{-l}$
and
$$
z'(w)=r\prod_{i=0}^{m-1}(1-a_iw^{-1})
$$
is a polynomial in $w^{-1}$
(all poles  $b_i$ of $z'(w)$
merge at the origin).
Then the
last sum in (\ref{PA11}) becomes
$m\log r$ and
the formula (\ref{PA11}) gives
\beq\label{eyn1}
F_{1}=-\frac{1}{24} \log \left (
r^4 \prod_{z'(a_j)=0}\frac{\bar z'(a_{j}^{-1})}{r}\right )=
-\frac{1}{24} \log \left (
r^4 \prod_{i,\,j=0}^{m-1}(1-\bar a_ia_j)\right )
\eeq
This formula is essentially identical  to the genus-1 correction
to the free energy of the Hermitian 2-matrix model
with a polynomial potential computed in \cite{Eynard1}.

\subsection{The determinant formula.}
For polynomial $V(z)$
$F_1$ enjoys an interesting
determinant representation.
Set
$$
D_m:=\det \left (\frac{\p^3 F_0}{\p t_{0} \p t_{j} \p t_{k}}
\right )_{0\leq j,k \leq m-1}
$$
where $F_0$ is the leading contribution to the free
energy regarded as a function of $t_0$
and the coefficients $t_k$
(and the complex conjugate coefficients $\bar t_k$).
We need the
{\it residue formula}
for the third order derivatives
of $F_0$ \cite{BMRWZ}:
\beq\label{res}
\frac{\p^3 F_0}{\p t_{j} \p t_{k} \p t_{l}}=
\frac{1}{2\pi i}\oint _{|w|=1}
\frac{h_j (w) h_{k} (w) h_l (w) }{z'(w)\bar z' (w^{-1})}
\, \frac{dw}{w}
\eeq
Here $h_j (w)$
is the following polynomial in $w$ of degree $j$:
$$
h_j (w)=w\frac{d}{d w}
\left [ (z^j (w))_{+} \right ]
\;\; \mbox{for $j\geq 1$ and}
\;\;\;h_0 (w)=1\,,
$$
where $(...)_+$ is the positive degree
part of the Laurent series.
Using this formula, we compute:
\beq\label{det2}
D_m= \frac{1}{(2\pi i)^m}\oint_{|w_0|=1}
\frac{dw_0}{w_0} \, \ldots \, \oint_{|w_{m-1}|=1}
\frac{dw_{m-1}}{w_{m-1}}\,\,
\frac{\mbox{det}\,
\left [ h_j (w_j )h_k (w_j)\right ]}{\prod_{l=0}^{m-1}
z'(w_l)\bar z'(w_{l}^{-1})}
\eeq
Clearly, the determinant in the numerator can be substituted
by $\frac{1}{m}\det^2 (h_j(w_k))$ and
$\mbox{det}\,
\left [ h_j (w_k )\right ]=
(m-1)! \, r^{\frac{1}{2}m(m-1)} \Delta_m (w_i)$,
where $\Delta_m (w_i)$ is the Vandermonde determinant.
Each integral in (\ref{det2}) is given by the sum of residues
at the points $a_i$ inside the unit circle (the residues at
$w_i =0$ vanish).
Computing the residues and summing over all permutations
of the points $a_i$, we get:
\beq\label{det3}
D_m=(-1)^{\frac{1}{2}m(m-1)}((m-1)!)^2 \,
r^{m(m-3)}
\frac{\prod_{j}a_{j}^{m-1}}{\prod_{i,\,j}(1-a_j\bar a_{i})}
\eeq
It is not difficult to see that
$\prod_{i=1}^{m}a_i = (-1)^m \, m(m\! -\! 1) r^{m-2}\bar t_m$
(we regard $t_m$, the last nonzero coefficient of the $V(z)$,
as a fixed parameter)
and so we represent $F_1$ (\ref{eyn1}) in the form
\beq\label{det4}
F_1 =\frac{1}{24}\log D_m
-\frac{1}{12}(m^2 \! -\! 3m\! +\! 3)\log r -
\frac{1}{24}(m\! -\! 1)\log \bar t_m +\mbox{const}
\eeq
where const is a numerical constant.
Since $\p^{2}_{t_0}F_0 =2\log r$ (see \cite{WZ2,Zles}), we
obtain that $F_1$, for models with polynomial potentials
of degree $m$,
is expressed through derivatives of $F_0$.

According to the conjecture of Section 4.1,
the  formula below
suggests a new representation of
the spectral determinant of the Laplace operator.
Up to  a constant we have
\beq\label{det5}
-\frac{1}{2} \log \det \left (- \Delta_{\Dc}\right )=
\frac{1}{24}\log\det_{m\times m}
\left (\frac{\p^3 F_0}{\p t_{0} \p t_{j} \p t_{k}}
\right )-\frac{1}{24}(m^2 \! -\! 3m\! +\! 3)
\, \frac{\p^2 F_0}{\p t_0^2}-
\frac{1}{24}(m\! -\! 1)\log \bar t_m
\eeq
where $j,k$ run from $0$ to $m-1$.
Similar determinant formulas are known
for genus-1 corrections to
free energy in topological field theories \cite{DZ}.
However, they
have not been identified
with spectral determinants.

\section{Spectral determinant of the
Laplace-Beltrami operator for the Dirichlet problem}

To give an interpretation of the result for $F_1$,
we recall the formula
for the spectral determinant of the
Laplace-Beltrami operator in a compact
bounded domain  $\MM$ in the plane
(assumed to have topology of a disk).
For the derivation see \cite{Polyakov,Alvarez},
Section 1 of \cite{OPS} and Appendix D.
The Laplace operator acts on functions vanishing
on the boundary (the Dirichlet boundary conditions).
 Being written in the Weyl gauge,
$g_{ab}=e^{2\Phi}\delta_{ab}$, it has the form
$$
\frac{1}{\sqrt{g}} \p_a ( \sqrt{g} g^{ab} \p_b )
=4e^{-2\Phi (w)} \p_w \p_{\bar w}=e^{-2\Phi}\Delta_{\MM}
$$
where $w$ is a holomorphic coordinate on $\MM$.
The Polyakov-Alvarez formula
 gives the difference between the spectral determinants
of the Laplace operators in
 the metric $e^{2\Phi}dw d\bar w$
and in some fixed
reference metric. Assuming that the latter
is just the standard flat
metric $dw d\bar w$ in the plane,
the formula reads (see (4.42)
in the first paper in \cite{Alvarez}):
\beq\label{PA5}
\begin{array}{lll}
\log\det (-e^{-2\Phi}\Delta_{\MM})&=&
\displaystyle{
-\frac{1}{4\pi \epsilon^2} \int_{\MM}e^{2\Phi}d^2 w +
\frac{1}{4\sqrt{\pi}\epsilon}\oint_{\p \MM} e^{\Phi}|dw|
+\frac{1}{6}\log \epsilon\, }
\\&&\\
&&\displaystyle{
-\frac{1}{12\pi} \int_{\MM} |\nabla \Phi|^2 d^2w
-\frac{1}{6\pi} \oint_{\p \MM} \hat \kappa \Phi
|dw| -\frac{1}{4\pi}\int_{\MM} \Delta \Phi d^2 w}
\end{array}
\eeq
Here $\epsilon$ is an ultraviolet  cutoff and
$\hat \kappa$ is the curvature of the boundary
w.r.t. the reference metric.
The first three terms diverge as
$\epsilon \to 0$.
A sketch of the derivation is given in
Appendix D.

We are going to show that our result for the
``quantum" part of the free energy,
$F_1^{(0)}$,
given by (\ref{F10}), (\ref{ff3})--(\ref{ff4})
 agrees with (\ref{PA5})
generalized to exterior domains. Indeed,
let us adopt this formula for the exterior of the domain $\DD$.
For this purpose, we map
it to the exterior of the unit circle $\Uc$
by the conformal map  $z\to w(z)$ and
choose $\MM$ in (\ref{PA5}) to be  the exterior of the unit circle,
$\MM = \Uc$, with the metric
$\Phi (w)=\chi (z(w))+\phi (w)$. Then
\beq\label{C1}
e^{2\Phi}dw d\bar w =
e^{2\chi} dz d\bar z
\eeq
Taking into account that for the exterior of the unit circle
$\hat \kappa =-1$, we rewrite (\ref{PA5}) as
\beq\label{C2}
\begin{array}{lll}
\log\det (-e^{-2\chi} \Delta_{\Dc})&=&
\displaystyle{
-\frac{1}{4 \epsilon^2} \int_{\Dc}\sigma \, d^2 z +
\frac{1}{4\epsilon}\oint_{\p \DD} \sqrt{\sigma} \, |dz|
+\frac{1}{6}\log \epsilon\, }
\\&&\\
&&\displaystyle{
-\frac{1}{12\pi} \left [
\int_{|w|>1} |\nabla \Phi |^2 d^2w
- 2 \oint_{|w|=1}  \Phi
|dw| \right ] -\frac{1}{4\pi}\int_{|w|>1} \Delta \Phi d^2 w}
\end{array}
\eeq
On the complex plane $\CC$ (without
boundary) with coordinate $z$ in the metric $e^{2\chi}$ we have:
\beq\label{C3}
\log\det (-e^{-2\chi}\Delta_{\CC})=
-\frac{1}{4 \epsilon^2} \int_{\CC}\sigma \, d^2 z
+\frac{1}{3}\log \epsilon\,
-\frac{1}{12\pi}
\int_{\CC} |\nabla \chi|^2 d^2z
\eeq
Therefore, we can write:
\beq\label{C4}
\begin{array}{c}
\displaystyle{
\log \frac{\det (-e^{-2\chi}\Delta_{\CC})}{\det
(-e^{-2\chi}\Delta_{\Dc})}}=
\displaystyle{
-\, \frac{t_0}{4\epsilon^2} +\frac{1}{6}\log \epsilon -
\frac{P(\DD)}{4\sqrt{\pi}\epsilon} \, }
\\ \\
\displaystyle{
+ \, \frac{1}{12\pi} \left [
\int_{|w|>1} |\nabla \Phi |^2 d^2w
- 2 \oint_{|w|=1}  \Phi \, |dw|
-3\oint_{|w|=1} \p_n \chi \, |dw| \right ] -
\frac{1}{12\pi}
\int_{\CC} |\nabla \chi|^2 d^2z}
\end{array}
\eeq
The first two divergent terms
can be absorbed by normalization and so we
ignore them in what follows. The third one is proportional
to the perimeter of the density support (in the metric
$e^{2\chi}$):
$$
P(\DD )=\oint_{\p \DD} \sqrt{\pi \sigma}\, ds
$$
It can be directly verified that the
perimeter functional obeys the relation
$$
\int \frac{\delta P(\DD )}{\delta W(z )}\,
L(a, z )\, d^2 z =0
\quad \quad (a\in \Dc )
$$
so all the three divergent terms
are ``zero modes'' of the loop operator
(for the first two this is obvious).

Comparing (\ref{C4}) with (\ref{ff3}), we can represent
$F_1$ in the form
\beq\label{C5}
F_1 =
\frac{1}{2}\, \log  \frac{\det'
(-e^{-2\chi}\Delta_{\CC})}{\det'
(-e^{-2\chi}\Delta_{\Dc})}\,
+ \, \frac{\alpha^2}{4\pi}\left [ \int _{\DD} |\nabla \chi |^2
d^2 z \! -\! \oint _{\p \DD}\chi \p_n \chi^H \, ds\right ]  +
\frac{\mu +\frac{1}{4}}{2\pi} \oint_{\p \DD}\p_n \chi  ds
\eeq
Here $\log \det '$ means the finite (as $\epsilon \to 0$)
part of (\ref{C4}). It is
the ``quantum" part of the answer.
Obviously, $F_1$ is related to gravitational
anomalies. This relation awaits further clarification.

\section{Appendices}

\subsection*{A. Some useful formulas}

Here we fix the notation and collect some formulas
often used in the main text.

\paragraph{Integral formulas.}
Throughout the paper, all contours are assumed
to be anticlockwise oriented and
the normal vector looks outward $\DD$.
\begin{itemize}
\item
Cauchy's integral formula ($f$ is any smooth function):
$$
\frac{1}{2\pi i}\oint_{\p \DD}
\frac{f(\zeta )d\zeta }{z-\zeta }
-\frac{1}{\pi}\int_{\DD}
 \frac{\p_{\bar \zeta}f(\zeta )d^2 \zeta}{z-\zeta }
=\left \{\begin{array}{cl}
-f(z)\,, & \;\; z\in \DD \\
0\,, &\;\; z\in \Dc
\end{array}\right.
$$
\item
The Green formula:
$\displaystyle{
\int_{\DD} f \Delta g d^2z = -\int_{\DD}
 \nabla f \, \nabla g \, d^2z
+\oint_{\p \DD}f \p_n g |dz|}
$.
\item
The Hadamard variational formula: the variation
of the Green function of $\Dc$ under small deformation of
the domain $\DD$ with the normal displacement $\delta n(z)$
is
$$
\delta G(z_1 , z_2)=\frac{1}{2\pi}
\oint_{\p \DD} \p_n G(z_1 , \xi ) \p_n G(z_2 , \xi )
\delta n (\xi ) \, |d\xi |
$$
\end{itemize}

\paragraph{Variation of contour integrals.}
Consider the contour integral of the general form
$$\oint_{\p \DD} F(f (z), \p_n f(z))ds
$$
where
$F$ is any fixed function.
Calculating the linear response
to the deformation of the contour
(described by the normal displacement
$\delta n(z)$),
one should vary all items in the integral independently
and add the results. There are four elements to be varied:
the support of the integral $\oint$, the $\p_n$, the line element $ds$
and the function $f$. By variation of the $\oint$ we mean
integration of the old function over the new contour. This
gives
$\oint \delta n \, \p_n F \,ds.$
The change of the slope of the normal vector results in
$\delta (\p_n ) =-\p_s (\delta n) \p_s.$
The rescaling of the line element gives
$\delta ds =\kappa \,\delta n ds$,
where $\kappa (z)$ is the local curvature of the boundary curve.
At last, we have to vary the function $f$
if it explicitly depends on the contour.
In particular, if this function is the harmonic
extension of a contour-independent function on the
plane, its variation on the boundary is given by
$$
\delta f^H (z) = \p_n (f(z)-f^H (z)) \delta n (z)\,,
\quad \quad z\in \p \DD
$$
This is an equivalent form of the Hadamard variational
formula. (For more details see e.g. \cite{WZ2,MWZ}, where
the Hadamard formula is extensively used.)

\paragraph{The curvature.}
The local curvature of the boundary curve
is defined as
$\kappa =d\theta /ds$,
where $\theta$ is the angle between the outward pointing
normal vector
to the curve and the $x$-axis.
The formula
\beq\label{curv2}
\kappa (z)=\p_n \log \left |\frac{w(z)}{w'(z)}\right |
\eeq
is an immediate consequence of the definition.
For practical calculations, we also need
the Laplace operator at boundary points in terms of normal ($\p_n$)
and tangential ($\p_s$) derivatives,
$$
\Delta = \p_n^2 +\p_s^2 +\kappa \p_n
$$
and the formula for the tangential derivative
of the curvature, $\kappa ' =\p_s \kappa$, through the boundary
value of the Schwarzian derivative
$
\{w;z\}=\frac{w'''}{w'}-\frac{3}{2}\left (\frac{w''}{w'}\right )^2
$:
\beq\label{curv4}
\kappa ' (z)={\cal I}m \, \left (
\nu^2 (z)\{w;z\}\right )
\eeq
($\nu$ is the complex unit normal vector (\ref{nuw})).
The variation of the curvature under small deformations
of the contour is
\beq\label{curv3}
\delta \kappa (z)=-(\p_{s}^2 +\kappa ^2(z))\delta n(z)
\eeq
Indeed,
$
\kappa \! +\! \delta \kappa =\frac{d(\theta +\delta \theta )}{ds+
\delta ds}
$,
where $\delta \theta =\p_s (\delta n)$ is the change of the slope
of the normal vector. The line element is rescaled as
$ds \to ds +\kappa \,\delta n ds$, so $\delta ds =\kappa \,\delta n ds$.
In the first order in $\delta n$ we then have
$\kappa +\delta \kappa = \kappa -\p_{s}^{2} (\delta n)
-\kappa^2 \delta n$
that is (\ref{curv3}).

\subsection*{B. Variations of the trial functionals}

Let us consider in detail
the most complicated case of the functional
$$
I^{(2)}=-\,
\frac{1}{2\pi} \oint_{|w|=1}(\phi \p_n \phi +2\phi )|dw|
$$
To vary this
functional, it is convenient to
pass to the $z$-plane and
introduce the function
\beq\label{q}
q(z)=\log \left | \frac{w'(z)}{w(z)}\right |
\eeq
It is harmonic in $\Dc$ with the logarithmic
singularity at infinity: $q(z) =-\log |z| +O(1/|z|)$ as
$z\to \infty$.
Obviously, $q^H (z)=\log |w'(z)| =q(z)+\log |w(z)|$.
In terms of this function, the curvature is
$\kappa (z) =-\p_n q(z)$, and
\beq\label{Alv}
I^{(2)}=
\frac{1}{2\pi}\oint_{\p \DD} (-q\p_n q +q e^q ) ds
\eeq
We would like to find the linear response of this
quantity to small deformations of the contour.

For clarity, let us deal with the two terms
in (\ref{Alv}) separately.
We apply the rules
given above and get, after some cancellations:
$$
\delta \oint q \p_n q \,ds =
\oint \delta n\, |\nabla q |^2 ds +
\oint ( \delta q \p_n q +q\, \p_n \delta q )\,ds
$$
where $|\nabla q|^2 = (\p_n q)^2 + (\p_s q)^2$.
Using the Green theorem and the behaviour of the function
$q$ at infinity, one can see that the contributions
of the two terms
in the second integral are the same. The variation
$\delta q$ at the boundary can be found by means of the
Hadamard formula. The result is:
\beq\label{var1}
\delta q=-\, e^{-q}\p_n \left (
e^{q}\, \delta n \right )^H
\eeq
Therefore,
$$
\delta \oint q \p_n q \,ds =
\oint \left [ |\nabla q|^2 -2e^q \p_n
\left ( e^{-q}\p_n q \right )^H \right ] \delta n \, ds
$$
where we used the Green formula again.
The harmonic continuation is achieved by means of
the identity $e^{-q}\p_n q = 2{\cal R}e \, (\frac{w}{w'}
\p_z q)$ whose r.h.s. is explicitly harmonic and thus
provides the desired harmonic continuation.
Next, a straightforward calculation shows that
$$
e^q \p_n
\left ( e^{-q}\p_n q \right )^H
=- e^q \p_s \left ( e^{-q}\p_s q \right )
=(\p_s q)^2 - \p_{s}^{2} q
$$
and we get
$$
\delta \oint q \p_n q \,ds =
\oint \left (2\p_{s}^{2}q -|\nabla q|^2
+2\kappa^2 \right ) \delta n \, ds
$$
The combination
$2\p_{s}^{2}q -|\nabla q|^2$ can be transformed as follows:
$$
2\p_{s}^{2}q -|\nabla q|^2 =
-2\Bigl [ \nu^2 (z)\, (\p_{z}^{2}q \!-\! (\p_z q)^2 ) \, +\,
\mbox{c. c.}\, \, \Bigr ]
$$
where
$2\, (\p_{z}^{2}q \!-\! (\p_z q)^2 )=
\{w;z\} +\frac{1}{2}\left (\frac{w'}{w}\right )^2$ and
$\{w;z\}$ is the Schwarzian derivative.
Combining these formulas, we get:
\beq\label{111}
2\p_{s}^{2}q -|\nabla q|^2 =
-\left ( \nu^2 \{w;z\} +\overline{\nu^2 \{w;z\}} +
|w'|^2 \right )
\eeq
Now to the second term in (\ref{Alv}). We have:
$$
\begin{array}{lll}
\displaystyle{
\delta \oint qe^{q}\, ds}&=&
\displaystyle{\oint \delta n
\p_n q e^{q} + \oint \left (e^q \delta q  +
qe^q\,\delta q \right )ds}\, =
\\&&\\
&=&\displaystyle{
\oint \delta n e^q \p_n (q-q^H)\, ds}\, =
\displaystyle{
-\oint \delta n \, e^{2q}\, ds}
\end{array}
$$
where we use the identity
$
\p_n \log |w(z)|=|w'(z)|
$
and take into account that $|w'|=e^q$ at the boundary.
Combining the results,
we obtain:
\beq\label{varI1}
\delta I^{(2)}=\frac{1}{2\pi}
\oint  \left (\nu^2 (z) \{w;z\} +\overline{\nu^2(z) \{w;z\}}
-2\kappa^2(z) \right ) \delta n(z) \,ds
\eeq

Variations of other trial functionals go in a similar
way. Here we just list the results:
$$
\delta \int_{\DD}|\nabla \chi |^2 d^2z =
\oint_{\p \DD}
|\nabla \chi |^2 \delta n \, ds +2 \int_{\DD} \nabla \chi
\nabla (\delta \chi ) \, d^2 z
$$
$$
\delta \oint_{\p \DD}\kappa \chi \, ds =
\oint_{\p \DD} (-\p_{s}^{2} \chi +\kappa \p_n \chi )
\delta n \, ds + \oint_{\p \DD} \kappa \, \delta \chi \, ds
$$
$$
\delta \oint_{\p \DD} \chi \p_n \chi^H \, ds
=\oint_{\p \DD}
\left ( |\nabla \chi |^2 -
\left ( \p_n (\chi - \chi^H)\right )^2 \right )
\delta n \, ds + 2
\oint_{\p \DD}
\delta \chi \, \p_n \chi^H \, ds
$$
$$
\delta \int_{\DD} \Delta \chi d^2 z =
\oint_{\p \DD}
\Delta \chi \, \delta n \, ds +
\oint_{\p \DD}
\p_n (\delta \chi )ds
$$
To express the right hand sides through the variation of
the potential $W$, one should plug the formulas
(\ref{deltan}), (\ref{chivar}).

\subsection*{C. The centrosymmetric potential at
$\beta =1$}

In the centrosymmetric case the potential $W$
does not depend on the angular coordinate in the plane.
We set $W (z)=W_{{\rm rad}} (|z|^2)$.
At $\beta =1$, the orthogonal polynomials technique
is applicable. The symmetry of the potential implies that
the orthogonal polynomials are simply monomials $z^n$.
Therefore,
the expression for the partition function
simplifies considerably:
$$
Z_N =\prod_{n=0}^{N-1} h_n
$$
where
$$
h_n = \int_{{\bf C}} |z|^{2n} e^{\frac{1}{\hbar}W}d^2z =
\pi \int_{0}^{\infty} x^n e^{\frac{1}{\hbar}W_{{\rm rad}}(x)}dx
$$
Introduce the function
\beq\label{axi1}
h(t)= \int_{0}^{\infty} e^{\frac{1}{\hbar}(
W_{{\rm rad}}(x) +t \log x )}dx
\eeq
then
\beq\label{axi2}
\log Z_N =N\log \pi +\sum_{n=0}^{N-1} \log h(n\hbar )
\eeq
We are going to find the $\hbar$-expansion of $\log Z_N$
using the $\hbar$-expansion
of $h(t)$ obtained by the saddle point method
and the Euler-MacLaurin formula. The necessary
formulas are collected below.

\paragraph{$\hbar$-Expansion of the free energy:
a direct calculation.}
First of all we note that the density support $\DD$ is
an axially symmetric domain. We assume that
it is a disk (not a ring) of radius $R$.
The density of eigenvalues is
$\sigma (z) =
\sigma_{{\rm rad}}(|z|^2)$.
Clearly,
$\pi \sigma_{{\rm rad}} (x) =
- \p_x (x \p_x W_{{\rm rad}} (x ))$.
The radius $R$
is a function of $t_0$ defined by the relation
\beq\label{ax1}
t_0 =-R^2 W_{{\rm rad}}'(R^2)
\eeq
(here and below prime means $f'(x)=df/dx$).
This relation follows directly from the definition
$$
t_0 =-\frac{1}{4\pi}\int_{\DD} \Delta W \, d^2 z =
-\frac{1}{4\pi}\oint_{\p \DD} \p_n W\, ds
$$
taking into account that $ \p_n W = 2R W_{{\rm rad}}'(R^2)$.

The saddle point $x_c$ in (\ref{axi1}) is determined by
the equation
\beq\label{axi3}
t=-x_c W_{{\rm rad}}' (x_c )
\eeq
Note the similarity with (\ref{ax1}): $x_c$ is the squared
radius of the disk for the filling $n=t/\hbar$.
We assume that there is only one saddle point.
The standard technique
(see (\ref{as4}))
yields the following asymptotic expansion as $\hbar \to 0$:
\beq\label{axi4}
h(t)=e^{\frac{1}{\hbar}(W_{{\rm rad}} (x_c )-
W_{{\rm rad}}' (x_c )x_c \log x_c )}
\sqrt{\frac{2 \hbar x_c}{\sigma_{{\rm rad}} (x_c )}}
\left ( 1+\hbar p_1 +O(\hbar^2) \right )
\eeq
where $p_1$ is given by (\ref{as5}) with $S(x)=W_{{\rm rad}} (x )+
t\log x$.
It is easy to calculate:
$$
\begin{array}{l}
\displaystyle{
S''(x_c) =-\frac{ \pi \sigma_{{\rm rad}} (x_c )}{x_c} }
\\ \\
\displaystyle{
S'''(x_c) =-\frac{ \pi \sigma_{{\rm rad}}' (x_c )}{x_c} +
2\frac{ \pi \sigma_{{\rm rad}} (x_c )}{x_c^2} }
\\ \\
\displaystyle{
S^{IV}(x_c) =-\frac{ \pi \sigma_{{\rm rad}}'' (x_c )}{x_c} +
3\frac{ \pi \sigma_{{\rm rad}}' (x_c )}{x_c^2} -
6\frac{\pi \sigma_{{\rm rad}} (x_c )}{x_c^3} }
\end{array}
$$
Plugging this stuff into (\ref{as5}), we get:
$$
p_1 = \frac{1}{24 \pi \sigma_{{\rm rad}} (x_c)}
\left ( 5x_c \left ( \frac{\sigma_{{\rm rad}}'
(x_c )}{\sigma_{{\rm rad}} (x_c )}\right )^2 -11
\frac{\sigma_{{\rm rad}}'
(x_c )}{\sigma_{{\rm rad}} (x_c )} -3 x_c
\frac{\sigma_{{\rm rad}}''
(x_c )}{\sigma_{{\rm rad}} (x_c )} +\frac{2}{x_c}\right )
$$
In terms of the function
$\chi (z)=\chi_{{\rm rad}} (|z|^2)=\frac{1}{2}\log
(\pi \sigma_{{\rm rad}})$
this expression acquires the form
$$
p_1 = \frac{1}{\pi \sigma_{{\rm rad}} (x_c)}\left (
\frac{1}{3}x_c (\chi_{{\rm rad}}' (x_c))^2
-\left. \frac{1}{4} \p_x (x\p_x \chi_{{\rm rad}} (x))
\right |_{x=x_c}\!\!
-\, \frac{2}{3} \chi_{{\rm rad}}' (x_c) +\frac{1}{12 x_c}\right )
$$

The Euler-MacLaurin formula applied to (\ref{axi2}) yields
$$
\log Z_N = N\log \pi +\frac{1}{\hbar}
\int_{0}^{t_0}\log h(t)dt -\frac{1}{2} \log h(t_0)
+\frac{\hbar}{12}\p_t \log h (t)\Bigr |_{t=t_0} + C
$$
where $C$ is a constant which does not depend on $t_0$.
(It can be found by a more detailed analysis around the point
$t=0$.)
From (\ref{axi4}) we have:
\begin{eqnarray}
\log h(t)&=&\frac{1}{\hbar}
(W_{{\rm rad}}(x_c)-W_{{\rm rad}}'(x_c)\, x_c \, \log x_c )
+\log \sqrt{2\pi \hbar} +
\nonumber\\
&+& \frac{1}{2} \log x_c -\chi_{{\rm rad}}(x_c) +\hbar p_1
+O(\hbar^2)
\end{eqnarray}
The integral over $t$ is transformed into the integral over
$x_c$ using $dt/dx_c =\pi \sigma_{{\rm rad}} (x_c)$:
$$
\int_{0}^{t_0}\log h(t)dt =
\pi \int_{0}^{R^2} \sigma_{{\rm rad}} (x)\log h(t(x))dx
$$
Finally, we obtain:
$$
F_0 =\pi \int_{0}^{R^2}\!
(W_{{\rm rad}}(x)-W_{{\rm rad}}'(x)\, x \, \log x )\,
\sigma_{{\rm rad}} (x)dx
$$
$$
F_{1/2} =-\, \frac{\pi }{2}\int_{0}^{R^2}\!
\sigma_{{\rm rad}} (x)\, \log (\pi \sigma_{{\rm rad}} (x)) dx
$$
\beq\label{F18}
F_{1} =-\frac{1}{12}\log R^2 -
\frac{1}{6}\chi_{{\rm rad}}(R^2)
-\frac{1}{4}R^2 \chi_{{\rm rad}}' (R^2)
+\frac{1}{3}\int_{0}^{R^2}x (\chi_{{\rm rad}}' (x))^2 dx
\eeq

Reducing the 2D integrals in (\ref{ff3}) to 1D integrals or values
of $\chi_{{\rm rad}}$, $\chi_{{\rm rad}}'$ at the boundary,
$$
\begin{array}{l}
\displaystyle{
\int_{\DD} |\nabla \chi |^2 d^2z = 4\pi \int_{0}^{R^2}
x(\chi_{{\rm rad}}' (x))^2 dx }
\\ \\
\displaystyle{
\oint_{\p \DD} \kappa \chi ds = 2\pi \chi_{{\rm rad}} (R^2)}
\\ \\
\displaystyle{
\int_{\DD} \Delta \chi \, d^2z = 4\pi R^2
\chi_{{\rm rad}}' (R^2 ) }
\end{array}
$$
and taking into account that $\p_n \chi^H =0$, we have:
$$
F_1 = -\frac{1}{12} \log R^2 -\frac{1}{6} \chi_{{\rm rad}} (R^2)
+\frac{1}{3}\int_{0}^{R^2}x(\chi_{{\rm rad}}' (x))^2 dx
+2\mu R^2 \chi_{{\rm rad}}' (R^2 )
$$
Comparing with (\ref{F18}), we conclude that
$\mu =-1/8$.

\paragraph{Asymptotic formulas.}
Evaluating the integral $\int e^{\frac{1}{\hbar}S(x)}dx$ by the
saddle point method around the critical point
$x_c$, $S'(x_c)=0$, we get the asymptotic expansion as $\hbar \to 0$:
\beq\label{as4}
\int e^{\frac{1}{\hbar}S(x)}dx =
 e^{\frac{1}{\hbar}S(x_c)}\sqrt{
\frac{2\pi \hbar}{|S''(x_c)|}} ( 1+ \hbar p_1 +\hbar^2 p_2 +\ldots )
\eeq
where
$$
p_n =\frac{\Gamma (n+\frac{1}{2})}{\sqrt{2\pi} (2n)!}
|S''(x_c)|
\left ( \frac{d}{dx}\right )^{2n}\left.
\left [ -\, \frac{S(x)-S(x_c)}{(x-x_c)^2} \right ]^{-n-\frac{1}{2}}
\right |_{x=x_c}
$$
In particular,
\beq\label{as5}
p_1 =
\frac{5(S'''(x_c))^2 +3|S''(x_c)|S^{IV}(x_c))}{24\,|S''(x_c)|^{3}}
\eeq
In the text we  also need the Euler-MacLaurin formula:
$$
\sum_{n=0}^{N-1} f(n) =
\int_{0}^{N} f(x)dx -
\frac{1}{2} ( f(N)- f(0))
+\frac{1}{12} ( f'(N)-f'(0)) +\ldots
$$

\subsection*{D. Derivation of the Polyakov-Alvarez
formula}

In this appendix we outline the derivation
of the formula (\ref{PA}) for the spectral determinant
of the Laplace operator. We use the notation
introduced in the main text.

Consider a free Bose field $X$ defined in a
planar (compact) domain  $\BB$
and  vanishing on its boundary.
For simplicity we assume that $\BB$ has
topology of a disk.
Let $e^{2\chi}dzd\bar z$ be
a background  conformal metric
on $\BB$.
The classical action
\beq
S=\frac{1}{4\pi}\int_{\BB}X(-e^{-2\chi} \Delta) X
\, e^{2\chi} d^2 z
\eeq
does not depend on the metric but
the quantum theory does.
The log of partition function  of this field
${\cal F}=-\frac{1}{2}\log\det
\left (-e^{-2\chi}\Delta_{\BB}\right) $
represents the spectral determinant of the
Laplace-Beltrami operator
$e^{-2\chi}\Delta_{\BB}=
4e^{-2\chi (z)} \p_z \p_{\bar z}$
acting on functions vanishing
on the boundary (the Dirichlet b.c.).
Let $z(w)$ be the univalent conformal map
from the unit disk $\UU$ onto $\BB$ and
$\phi =\log |dz/dw|$. This map induces
the conformal metric $e^{2\Phi}dw d\bar w$
on $\UU$ in the
holomorphic coordinate $w$, with the conformal factor
$$
e^{2\Phi} =
e^{2\chi}|z'(w)|^2 =
e^{2(\chi +\phi )}
=\sqrt{g}
$$
The Laplace-Beltrami operator in the
coordinate $w$ is $
4e^{-2\Phi (w)} \p_w \p_{\bar w}$.
The spectral  determinant of this operator
can be expressed in terms of $\Phi$
and the curvature of the boundary.
 The result is  referred to as
the Polyakov-Alvarez formula \cite{Polyakov,Alvarez}.
Below we give a short derivation of this formula.

Variation of the partition function
over the metric introduces the trace
$T_{z\bar z}=
\lbracket |\p X|^2\rbracket$
and the  holomorphic component
$T=\lbracket (\p X)^2\rbracket$
of the stress energy tensor:
\beq\label{911}
-\, \frac{1}{\pi}T_{z\bar z}=
\frac{1}{2}\left ( \frac{\delta}{\delta \chi}
+\frac{\delta }{\delta \phi}\right ){\cal F}=
\sqrt{g}\, \frac{\delta {\cal F}}{\delta \sqrt{g}}
\eeq
The trace of the s.e.t.
has bulk and boundary parts, and so has
${\cal F}$:
\beq
T_{z\bar z}dz d\bar z=
T_{z\bar z}^{\rm bulk}dz d \bar z+T_{z\bar z}^{\rm boundary}ds,
\quad
 {\cal F}={\cal F}^{\rm bulk}+{\cal F}^{\rm boundary}
\eeq
The holomorphic component is continuous
across  the boundary in a regular fashion.
The components $T_{z\bar z}$ and $T$
are related by the  conservation law
\beq\label{931}
\p (\sqrt{g}T_{z\bar z})+
\sqrt{g}\,\bar \p T =0 \quad  \mbox{(in the bulk)}
\eeq
\beq\label{93}
2\p_s T^{\rm boundary}_{z\bar z}+
{\cal I}m(\nu^2 T) =0 \quad \mbox{(on the boundary)}
\eeq
where $\nu$ is the unit normal vector to the boundary.

The strategy to compute ${\cal F}$  is as follows.
First we compute  the holomorphic
component $T$, then find $T_{z\bar z}$ through the
conservation laws (\ref{931}), (\ref{93}).
Finally we integrate eq. (\ref{911}).

The simplest way to proceed is as follows.
Since the holomorphic component is continuous across
the boundary, it depends only on the overall
 metric $\sqrt{g}=e^{2\chi + 2\phi}$.
Therefore, it is sufficient to
compute $T$ at $\chi =0$.
In the coordinate $w$, the domain is the unit disk,
so $T(w)(dw)^2$ vanishes.
The conformal transformation $w\to z$  generates the metric
$e^{-2\phi}$ and
transforms the holomorphic component of the stress energy tensor as
$T\to T+\frac{1}{12}\{w;z\}$. Therefore\footnote{One can obtain
this textbook result by the direct computation of
$\lbracket \partial X(z)\partial X(z')\rbracket
\propto \p_z \p_{z'} G(z,z')$ and extracting
the finite part of the result as $z'\to z$
with the help of the explicit
formula for the Green function (\ref{Gw}).},
$$
T(z)=\frac{1}{12}\{w;z\}\,
=-\frac{1}{24}\left [ \left (\p \log \left |
dz/dw\right |^2 \right )^2 + 2\p^2 \log
\left |dz/dw\right |^2 \right ]
$$
at $\chi =0$,
or, more generally,
\beq\label{}T=-\frac{1}{24}\left ((\p\log \sqrt{g})^2+
2 \p^2 \log \sqrt{g}\right )
\eeq
at $\chi \neq 0$.
The next step is to find the trace of
the stress energy tensor. The bulk part
\beq
T^{\rm bulk}_{z\bar z} =
\frac{1}{48}\Delta\log
\sqrt{g}= \frac{1}{24}\Delta \chi
\eeq
follows from (\ref{931}) and
represents the gravitational anomaly.
It is proportional
to the scalar curvature of the metric in the bulk
and does not depend on the shape of the domain.
This gives (after integrating eq. (\ref{911}))
the textbook result for the bulk part of
the spectral determinant:
\beq\label{92}
{\cal F}^{\rm bulk}=-\, \frac{1}{96\pi}
\int _{\UU }\log \sqrt{g} \, \Delta \log \sqrt{g}
\, d^2 w
\eeq

We will
use the same trick in order to compute the boundary part.
Let us first find it at $\chi =0$.
Comparing (\ref{93}) and (\ref{curv4}) we
conclude that it is
$\frac{1}{24}\p_s\kappa$, where $\kappa$
is the local curvature of the boundary.
This gives the boundary part of the gravitational anomaly:
$T_{z\bar z}^{\rm boundary}=-\frac{1}{24}\kappa$.
At $\chi \neq 0$ one should simply add
the boundary curvature generated by the metric
$e^{2\chi}$ to the local curvature $\kappa$:
\beq
T_{z\bar z}^{\rm boundary}ds=
-\, \frac{1}{24} (\kappa+\p_n \chi)ds\, .
\eeq
Using the formula $\kappa ds
=( \p_n\phi +e^{-\phi}) ds= \p_n\phi ds + |dw|$
(equivalent to (\ref{curv2})),
the boundary contribution to the gravitational anomaly
can be written as
$$
T_{z\bar z}^{\rm boundary}ds=
-\, \frac{1}{48} \left (2|dw|
+\p_n\log \sqrt{g}\, ds \right )
$$
This form is ready for integration with the result
\beq
 {\cal F}^{\rm boundary}=
\frac{1}{96\pi}\oint_{\p \UU}
\log \sqrt{g}\, \p_n\log \sqrt{g} \, |dw|+
\frac{1}{24\pi}\oint_{\p \UU}
\log \sqrt{g}\, |dw|+
\frac{1}{16\pi}\oint_{\p \UU} \p_n\log \sqrt{g}\,|dw|
\eeq
where we take into account that $\p_n$
on $\p \DD$ is $e^{-\phi}\p_n$ on the unit circle.
Combining with (\ref{92}), we
obtain the finite part of (\ref{PA})
(recall that $\log \sqrt{g}= 2\Phi$).

\section*{Acknowledgments}

We thank the organizers of the CRM Program
``Random Matrices, Random Processes
and Integrable Systems", where these
results were reported, for the kind hospitality
in Montreal and especially J.Harnad for the suggestion
to write this paper.
We are grateful to Al.Abanov, L.Che\-khov,
B.Dub\-ro\-vin, V.Ka\-za\-kov and I.Kos\-tov
for useful discussions.
P.W. was supported by the NSF MRSEC Program under
DMR-0213745 and NSF DMR-0220198.
The work of A.Z. was supported in part by RFBR grant 04-01-00642,
by grant for support of scientific schools
NSh-1999.2003.2 and by grant NWO 047.017.015.

\end{document}